\documentclass[preprint,12pt]{elsarticle}

\usepackage{graphicx}

\usepackage{amssymb}

\biboptions{comma,square}

\journal{International Journal of Non-Linear Mechanics}

\begin{document}

\begin{frontmatter}



\title{Testing a Fast Dynamical Indicator: The MEGNO}


\author[L1,L2]{N.P. Maffione\corref{cor1}}
\author[L1,L2]{C.M. Giordano}
\author[L1,L2]{P.M. Cincotta}

\address[L1]{Facultad de Ciencias Astron\'omicas y Geof\'isicas, Universidad Nacional de La Plata, Paseo del Bosque, B1900FWA La Plata, Argentina}
\address[L2]{Instituto de Astrof\'isica de La Plata (CONICET, CCT-La Plata), Paseo del Bosque, B1900FWA La Plata, Argentina}

\cortext[cor1]{Corresponding author.\\ 
E-mail addresses: nmaffione@fcaglp.unlp.edu.ar (N.P. Maffione), pmc@fcaglp.unlp.edu.ar (P.M. Cincotta), giordano@fcaglp.unlp.edu.ar (C.M. Giordano)}

\begin{abstract}

To investigate non-linear dynamical systems, like for instance 
artificial satellites, Solar System, exoplanets or galactic models,
it is necessary to have at hand several tools, such as a reliable dynamical indicator.

\indent
The aim of the present work is to test a relatively
new fast indicator, the Mean Exponential Growth factor of Nearby Orbits (MEGNO), since it is becoming a widespread 
technique for the study of Hamiltonian systems, particularly in the field of dynamical astronomy and astrodynamics,
as well as molecular dynamics.

\indent
In order to perform this test we make a detailed numerical and statistical study of a sample of orbits in a triaxial galactic system, whose dynamics was investigated by means of the computation of the Finite Time Lyapunov Characteristic Numbers (FT--LCNs) by other authors.

\end{abstract}


\end{frontmatter}



\section{Introduction}
\label{S1}

In the present work we accomplish an exhaustive study of the MEGNO when applied
to a given sample of orbits in a triaxial galactic potential studied by \cite{MCW05}.
In that work, the authors use a well--known tool, the Lyapunov Characteristic Numbers (see e.g. \cite{S10}), to identify
the character of the selected orbits in order to classify them as regular or chaotic.

The MEGNO is introduced by \cite{CS00} and, in \cite{CGS03},
this technique is formalized and its application extended to discrete Hamiltonian
systems like maps; also a generalization of the MEGNO is introduced therein.
This tool has become of widespread use for studying several astronomical problems 
as well as many other Hamiltonian systems (see, e.g., 
\cite{G02,G03a,G03b,PM03,BKRP03,G04,GKW05,BMBW05,HMJGM08,GB08,LDV09,BBS09, HCA09, VDLC09}).

In \cite{CGS03} and \cite{GC04}, the MEGNO succeed in furnishing a clear
insight of the global structure of the phase space of simple multidimensional Hamiltonian systems,
providing a clear picture of the resonant network as well as the regular and chaotic domains.

Herein instead, a far more complex non--linear system is addressed that reproduces many
characteristics of real elliptical galaxies, namely, the one introduced by \cite{MCW05}.
This model will be used as the scenario for a detailed comparison between the MEGNO and the 
Lyapunov Characteristic Numbers and even the Fast Lyapunov Indicator (FLI) introduced by \cite{FGL97}.

There are many efficient dynamical indicators, some of them based on deviation vector(s), 
for instance, the spectra of stretching numbers, helicity and twist angles, the computation of the alignment indices introduced by Skokos, the Relative finite time Lyapunov Indicator (RLI) and also the Average Power Law Exponent (APLE), a technique recently developed \cite{CV97,S01,SBE00,LVE08}, and
others based on spectral analysis, such as  \cite{BS82}, the Frecuency Map Analysis~\cite{L90, L93},
the one due to Sidlichovsk\'y and Nesvorn\'y~\cite{SM96}, and perhaps the
latest one, the FMI (Frequency modulation indicator)~\cite{CO08}.  
However the present paper is devoted to accomplish a thorough test of  the 
MEGNO, pointing out not only its advantages but its drawbacks as well.
Therefore herein we just focus our attention on an exhaustive comparison of the MEGNO
against the Lyapunov Characteristic Numbers (and eventualy the FLI), since without any doubt, the latter
is the most widespread tool  in, at least, the last forty years, and it is still being used by many authors.
On the other hand, as far as we know, a full test of the MEGNO when applied to  a non--linear somewhat realistic Hamiltonian
system has not been performed yet.

\section{The Mean Exponential Growth factor of Nearby Orbits~(MEGNO)}
\label{S2}

In this section we summarize the main features of the MEGNO (described in detail in \cite{CGS03}). 
This is an alternative tool to explore the phase space which belongs to the class of the so--called 
fast indicators.

Let $H({\mathbf{p}},{\mathbf{q}})$ with ${\mathbf{p}},\,{\mathbf{q}}\in \mathbb{R}^N$
be an $N$--dimensional Hamiltonian, that we suppose autonomous just for the sake of simplicity. 
Introducing the following notation:

$${\mathbf{x}}=({\mathbf{p}},{\mathbf{q}})\in
\mathbb{R}^{2N},\, {\mathbf{v}}=(-\partial H/\partial{\mathbf{q}},\ \partial H/\partial
{\mathbf{p}})\in\mathbb{R}^{2N},$$

\noindent
the equations of motion can be written in a simple way like 

\begin{equation}
\dot{\mathbf{x}}={\mathbf{v}}({\mathbf{x}}).  
\label{eq2.1}
\end{equation}

Let $\gamma(\mathbf{x_{0}};t)$
be an arc of an orbit of the flow (\ref{eq2.1})
over a compact energy surface: $M_h\subset\mathbb{R}^{2N}$,
$M_h=\{{\mathbf{x}}: H({\mathbf{p}},{\mathbf{q}})=h\}$ with $h=$ $constant$,
then
$$
\gamma(\mathbf{x_{0}};t)=\{{\mathbf{x}}(t';{\mathbf{x}}_0):{\mathbf{x}}_0\in M_h,\ 0\le t'< t\}.
$$

We can gain fundamental information about the Hamiltonian flow in the neighborhood of any orbit $\gamma$ through 
the largest Lyapunov Characteristic Number (LCN) defined as:
\begin{equation}
\sigma(\gamma)=\lim_{t\to\infty}\sigma_1(\gamma(\mathbf{x_0};t)),\quad \sigma_1(\gamma(\mathbf{x_0};t))=
{1\over t}\ln\left[\|\vec{\delta}\gamma(\mathbf{x_0};t)\|\right], 
\label{eq2.2}
\end{equation}
with $\vec{\delta}\gamma(\mathbf{x_0};t)$ an ``infinitesimal displacement'' from $\gamma$ at time $t$, 
where $\|\cdot\|$ is some norm.
The fact that the LCN measures the mean exponential rate of divergence of nearby orbits it is clearly understood when Eq. (\ref{eq2.2}) is written in an integral fashion:
\begin{equation}
\sigma(\gamma)=\lim_{t\to\infty}{1\over t}\int_0^t{\dot{\delta}\gamma(\mathbf{x_0};t')
\over\delta\gamma(\mathbf{x_0};t')}{\rm d}t'= \overline{\left(\dot{\delta}/\delta\right)},
\label{eq2.3}
\end{equation}
where $\delta\equiv\|\vec{\delta}\|,\,\dot{\delta}\equiv {\rm d}\delta/{\rm d}t=
\dot{\vec{\delta}}\cdot\vec{\delta}/\|\vec{\delta}\|$, and the bar denotes time average. Also, the tangent vector $\vec{\delta}$ satisfies the variational equation
$$
\dot{\vec{\delta}}=\Lambda(\gamma(\mathbf{x_0};t))\cdot\vec{\delta},
$$
where $\Lambda$ is the Jacobian matrix associated with the vector field ${\mathbf{v}}$.

Now we are in a position to introduce the MEGNO,  $Y(\gamma(\mathbf{x_0};t))$, through the expression:
$$
Y(\gamma(\mathbf{x_0};t))={2\over t}\int_0^t{\dot{\delta}\gamma(\mathbf{x_0};t')\over
\delta\gamma(\mathbf{x_0};t')}t'{\rm d}t',
$$
which is related with the integral in Eq. (\ref{eq2.3});
i.e., in case of an exponential increase of $\delta$, $\delta\gamma(\mathbf{x_0};t)=
\delta\gamma(\mathbf{x_0};t_0)\cdot
\exp(\lambda t)$, the quantity $Y(\gamma(\mathbf{x_0};t))$ can be considered as a weighted 
variant of the integral in Eq. (\ref{eq2.3}).
Instead of using the instantaneous rate of increase, $\lambda$,
we average the logarithm of the growth factor, 
$\ln\left[\delta\gamma(\mathbf{x_0};t)/\delta\gamma(\mathbf{x_0};t_0)\right]=\lambda t$.

Let us describe the MEGNO's asymptotic behavior to exhibit its ability to give a clean idea of the character of orbits.
Firstly, consider the case of orbits on irrational tori for a non--isochronous system.
As it is shown in \cite{CGS03},
for quasi--periodic orbits, $\gamma_{q}$,
the temporal evolution of $Y(\gamma_q(\mathbf{x_0};t))$ is given by
$$
Y\left(\gamma_q(\mathbf{x_0};t)\right)\approx 2-{\ln(1+\lambda_{q}\,t)^{2}\over\lambda_{q}
\,t}+ O\left(\gamma_q(\mathbf{x_0};t)\right),
$$
where $\lambda_{q}$ is the linear rate of divergence around $\gamma_{q}$ and
$O$ is a null average oscillating term.
Accordingly to this formula, the $\lim_{t\to\infty}Y\left(\gamma_q(\mathbf{x_0};t)\right)$ 
does not exist, but on 
introducing a time average
$$
\overline{Y}(\gamma_q(\mathbf{x_0};t))\equiv{1\over t}\int_0^tY(\gamma_q(\mathbf{x_0};t')){\rm d}t',
$$
it can be found that
$$
\overline{Y}\left(\gamma_{q}\right)\equiv\lim_{t\to\infty}\overline{Y}
\left(\gamma_q(\mathbf{x_0};t)\right)=2.
$$
Then, for quasi--periodic motion, $\overline{Y}(\gamma)$
is a fixed constant, independent of $\gamma$.

When taking irregular orbits $\gamma_i$, i.e. orbits on some stochastic layer, for which 
$\delta\gamma_i(\mathbf{x_0};t)\approx\delta\gamma(\mathbf{x_0};t_0)\cdot \exp{(\sigma_it)}$, 
$\sigma_i$ being the LCN of $\gamma_i$, the temporal evolution of the MEGNO
will be given by:
$$
Y\left(\gamma_i(\mathbf{x_0};t)\right)\approx\sigma_it+\tilde{O}\left(\gamma_i(\mathbf{x_0};t)\right),
$$
with $\tilde{O}$ some bounded amplitude and null average oscillating term (see \cite{CGS03}). 
On averaging over a sufficiently large interval we have:
$$
\overline{Y}\left(\gamma_i(\mathbf{x_0};t)\right)\approx{\sigma_i\over 2}\,t,\quad t\to
\infty.
$$
Therefore, in the case of chaotic orbits, not only $Y\left(\gamma_i(\mathbf{x_0};t)\right)$ but also
$\overline{Y}\left(\gamma_i(\mathbf{x_0};t)\right)$ grow linearly with time, with a slope equal to the 
LCN of the orbit 
or one half of it, respectively. 
Wherever the phase space has a hyperbolic structure, $\overline{Y}$ will indefinitely grow with time. 
Otherwise, it will approach a constant value, even in the degenerated case in which $\delta$ grows 
with some power of $t$, e.g. $n$, for which $\overline{Y}\to 2n$ when $t\to\infty$.

We notice that the temporal evolution of the MEGNO can be briefly described in a suitable and unique 
expression for all kind of motion.
Indeed, the asymptotic behavior of $\overline{Y}(\gamma(\mathbf{x_0};t))$
can be summarized in the following way:
$\overline{Y}(\gamma(\mathbf{x_0};t))\approx a_{\gamma}t+d_{\gamma}$, where
$a_{\gamma}=\sigma_{\gamma}/2$ and $d_{\gamma}\approx 0$ for irregular, stochastic motion, 
while $a_{\gamma}=0$ and $d_{\gamma}\approx 2$ for quasi--periodic motion.
Deviations from the value $d_{\gamma}\approx 2$ indicate that $\gamma$
is close to some particular objects in phase space, being $d_{\gamma}\lesssim 2$ 
or $d_{\gamma}\gtrsim 2$ for stable periodic orbits (or resonant elliptic tori), 
or unstable periodic orbits (or hyperbolic tori) respectively (see \cite{CGS03} for details).
Finally, the quantity $\hat{\sigma}_1=Y/t$ verifies that
$$
\hat{\sigma}_1(\gamma_q(\mathbf{x_0};t))\approx{2\over t},\qquad\hat{\sigma}_1(\gamma_i(\mathbf{x_0};t))
\approx\sigma_i,\qquad as\qquad t\to\infty,
$$
supporting the fact that in regular domains, $\hat{\sigma}_1$ converges to 0
faster than $\sigma_1$ (which goes to zero like $\ln t/t$),
while for stochastic domains, both quantities tend to the positive LCN at a rather similar rate.

Let us introduce here a brief comment regarding the computation of
the LCN.
As it is already well--known, though the definition of the Lyapunov Characteristic Numbers encompasses
an integration over an infinite interval of time, their numerical computation involve
a rather large but finite time interval and the expected null value corresponding to
regular motion is unlikely to be reached. In such a case instead, the 'Finite
Time Lyapunov Characteristic Numbers' (FT--LCNs hereafter, following the nomenclature
given in \cite{VKS02})
attains a value of order $\ln{T}/T$,  being $T$ the total integration time.
Thus, a critical value has to be adopted as 'zero', so that
FT--LCNs' values greater or lower than such critical value are regarded as
different from or equal to zero respectively.

\section{The potential}
\label{S3}
 
For the comparison of the MEGNO vs. the FT--LCNs
 we deal with the potential introduced by \cite{MCW05} which,
obtained after the virialization of an $N$--body self--consistent model
composed of one hundred thousand particles, reproduces many features of
real elliptical galaxies, such as mass distribution,
flattening, triaxiality and rotation (see also, \cite{M06}).
Nonetheless, it is clear that a real elliptical galaxy is a much
more complex astrophysical system than a purely dynamical one.

This potential seems to provide an adequate scenario for the comparison 
between the two above mentioned techniques. To this aim, we address 
the study of the set of randomly selected orbits 
$\mathbf{O}=\{\mathbf{x}_i(t), i=1,\cdots, 3472,\ \mathbf{x}_i(0)=\mathbf{{x_0}}_i\}$, 
classified by means of the FT-LCNs in \cite{MCW05}. 
Their initial conditions $\mathbf{{x_0}}_i$ and their concomitant FT--LCNs values were
provided by the authors. We will identify each orbit of the set $\mathbf{O}$ with a label that runs from
1 to 3472.

The equation that reproduces the potential is:

\begin{equation}V(x,y,z)=-f_{0}(x,y,z)-f_{x}(x,y,z)\cdot(x^{2}-y^{2})-f_{z}(x,y,z)\cdot(z^{2}-y^{2}),\label{eq3.1}
\end{equation}
where
\begin{equation}
f_{n}(x,y,z)=\frac{\alpha_{n}}{\left[p_{n}^{a_{n}}+\delta_{n}^{a_{n}}\right]^{\frac{ac_{n}}{a_{n}}}},
\label{eq3.2}
\end{equation}
\\

\noindent
where $p^{2}_{n}$ is the square of the softened radius given by  
$p_{n}^2=x^2+y^2+z^2+\epsilon^2$ when $n=0$, or 
$p_{n}^2=x^2+y^2+z^2+2\cdot\epsilon^2$ for $n=x,z$, and 
$\alpha_{n}$, $\delta_{n}$, $a_{n}$, $ac_{n}$ are constants. 
The adopted value for the softening parameter is 
$\epsilon\simeq 0.01$ for any $n$. The functions $f_{n}(x,y,z)$ were computed 
through a quadrupolar $N$--body code for a hundred thousand bodies, which allowed 
the authors to write them in a general fashion given by Eq. (\ref{eq3.2}). 
The adopted values for the constants $\alpha_{n}$, $\delta_{n}$, $a_{n}$ and $ac_{n}$ 
are given in Table \ref{table 3.1} (further references in \cite{CGM08}).

\begin{table}[!ht]\centering
\begin{tabular}{ccccc}
\hline\hline  \vspace*{-2ex} \\ %
\emph{}  & $\alpha$ & $a$ & $\delta$ & $ac$\vspace*{1ex} \\ %
\hline 
$n=0$ & $0.92012657$ & $1.15$ & $0.1340$  & $1.03766579$\vspace*{1ex} \\%
\hline  %
$n=x$ & $0.08526504$ & $0.97$ & $0.1283$ & $ 4.61571581$\vspace*{1ex} \\%
\hline %
$n=z$ & $-0.05871011$ & $1.05$ & $0.1239$ & $4.42030943$\vspace*{1ex} \\%
\hline\hline  \vspace*{-4ex} %
\end{tabular}
\vspace{3mm}
\caption{Adopted values for the coefficients of the functions $f_n$ given by Eq. (\ref{eq3.2}).}
\label{table 3.1}
\end{table}

The stationary character of the parameters given in Table \ref{table 3.1} were  
tested by performing several fits at different times after virialization, 
resulting with a precision of $0.1\%$.

After the system had relaxed, there remained 86.818 bodies resembling 
an elliptical galaxy (the system obeying a de Vaucouleurs law, as Fig. 2 in 
\cite{MCW05} shows) with a strong triaxiality and a flattening that increases from the 
border of the system to its center (see Table I in the same work). 

\begin{figure}[!ht]
\begin{center}
\resizebox{65mm}{!}{\includegraphics{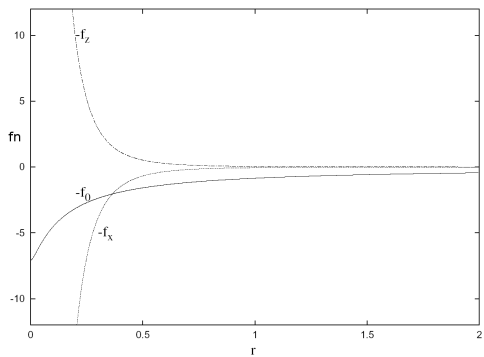}}
\caption{Radial dependency of the functions $f_{n}$ with their concomitant signs (figure taken from 
\cite{CGM08}.}
\label{figure 3.1}
\end{center}
\end{figure}

Fig. \ref{figure 3.1}, taken from \cite{CGM08}, displays the behavior of 
the $f_n$ regards to $r$,
being $f_z<0,\,\,\, f_0,\,f_x>0$, and $f_x > |f_z|$ for the whole $r$ range, 
while for $r\gtrsim 0.36$ it is $f_0 > f_x$. Notice that the functions $f_n$ are 
plotted with the concomitant sign with which they appear in Eq. (\ref{eq3.1}). 
The obtained triaxial potential has semi--axis $X,Y,Z$ satisfying   
the condition $X>Y>Z$ and its minimum, which is close to $-7$, matches the origin. 
The potential is less flattened than the mass distribution, as expected 
(see Table I in \cite{MCW05}).

\section{Comparison of the MEGNO vs. the FT--LCNs}
\label{S4}

\begin{figure}
\begin{center}
\resizebox{65mm}{!}{\includegraphics{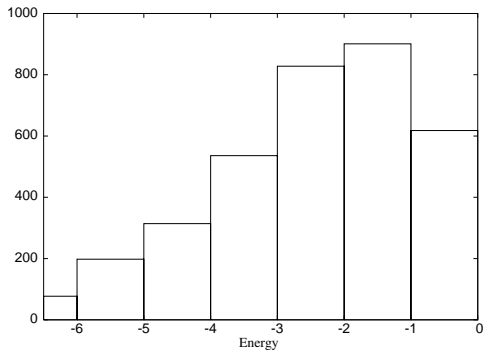}}
\caption{Energy spectrum of the 3472 orbits.}
\label{figure 4.1}
\end{center}
\end{figure}

The present section is devoted to performing a numerical and statistical comparative study of 
the results obtained by recourse to the MEGNO when applied to the set $\mathbf{O}$ of orbits and 
those provided by \cite{MCW05} through the FT-LCNs.

It is of interest to deem the energy spectrum of the orbits in the set $\mathbf{O}$ displayed in 
Fig. \ref{figure 4.1}, where we observe that most of the 3472 orbits considered have large energies;  
indeed, in the main they have energies in the range $-3\le E< 0$.

Let us recall the criterion used in \cite{MCW05} to classify the orbits in $\mathbf{O}$
according to their FT--LCNs: those orbits with their largest FT--LCN below some critical value $V_c$
were labeled as regular, otherwise they were classified as chaotic.

Since the numerical integrations carried out by \cite{MCW05} for the computation of the FT--LCNs 
encompasses an interval of 10000 u.t. (units of time), 
the expected value for $V_c$ would be $V_c^t=\ln T/T\approx 0.00092$ (u.t.)$^{-1}$.
Notwithstanding, \cite{MCW05} took an empirical value slightly higher, $V_c^e=0.00155$ (u.t.)$^{-1}$, 
and this is the one we consider in order to observe their classification into 
regular and irregular orbits. Along this investigation we adopt for the MEGNO 
a threshold value of $2.01$ for regular orbits.
According to \cite{MCW05}, the set $\mathbf{O}=\mathbf{Oc}\cup\mathbf{Or}$,  where 
$\mathbf{Oc}$ and $\mathbf{Or}$ include 1828 chaotic orbits and 1644 regular ones, respectively.

The computation of the MEGNO, as well as that of the largest FT--LCN, requires the integration of the equations of 
motion along with their first variationals, the initial conditions for the latter being taken at 
random in phase space and with unit norm. 
The integrations were accomplished using a \textit{Runge-Kutta 7/8 th} order integrator  
(the so--called \textit{DOPRI8} routine --see \cite{HNW87,PD81}--), 
over short:  5000 u.t,  intermediate: 10000 u.t. and large integration times: 100000 u.t. 
The precision in the conservation of the energy was of the order of $\sim 10^{-12}$.

The FT--LCN values corresponding to a total integration time of 10000 u.t. for the sample of
orbits classified in \cite{MCW05} were kindly provided by Muzzio, to whom we are grateful.

Herein we present the results corresponding to 5000 u.t. and to 10000 u.t. and even larger
motion times, in order to disclose how
efficient could this tool be, to provide dynamical information at short times.

\subsection{Results at short integration times}
\label{S4a}

\begin{figure}
\begin{center}
\begin{tabular}{cc}
\hspace{-20mm}\resizebox{90mm}{!}{\includegraphics{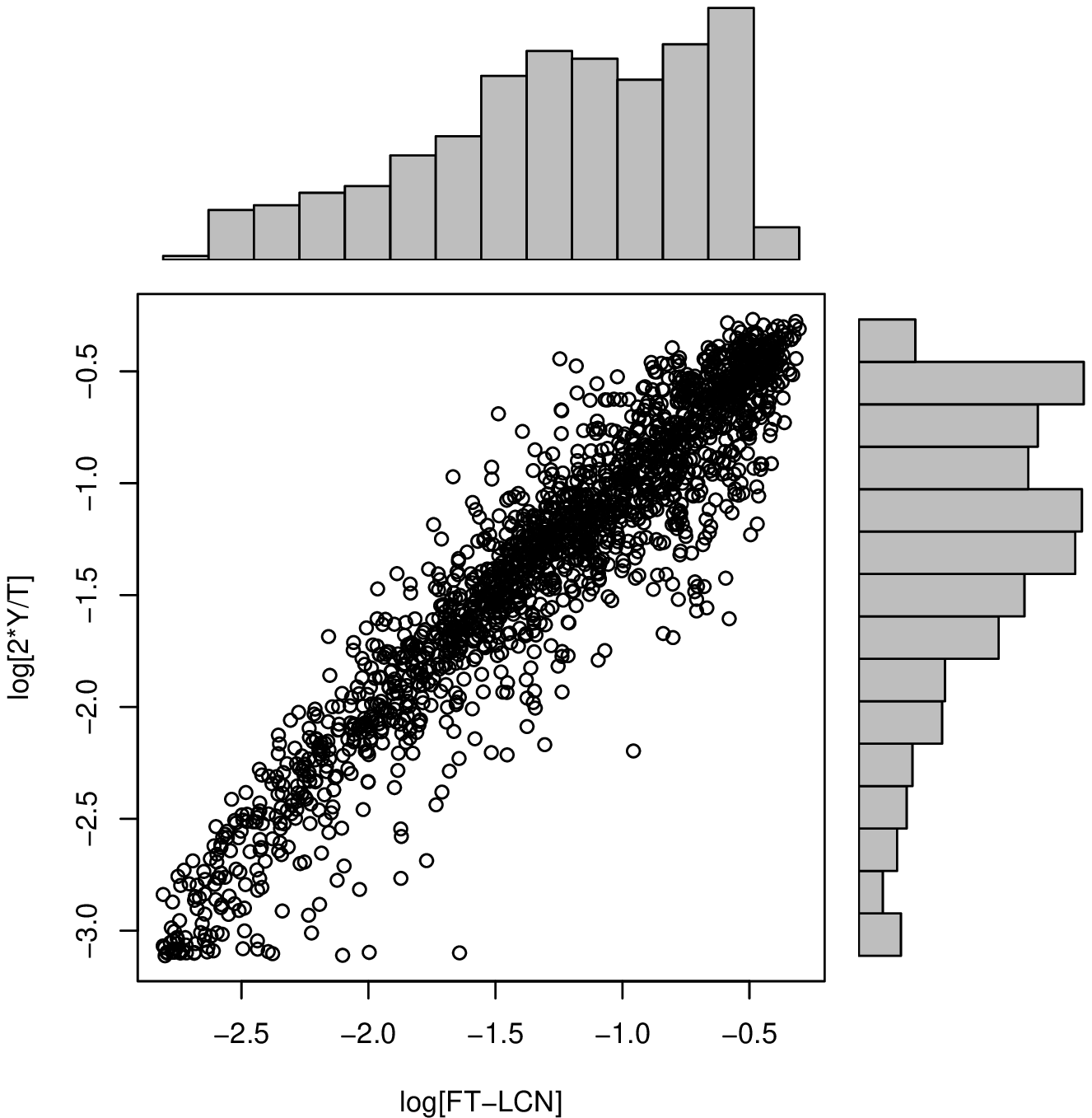}}&
\hspace{-30mm}\resizebox{90mm}{!}{\includegraphics{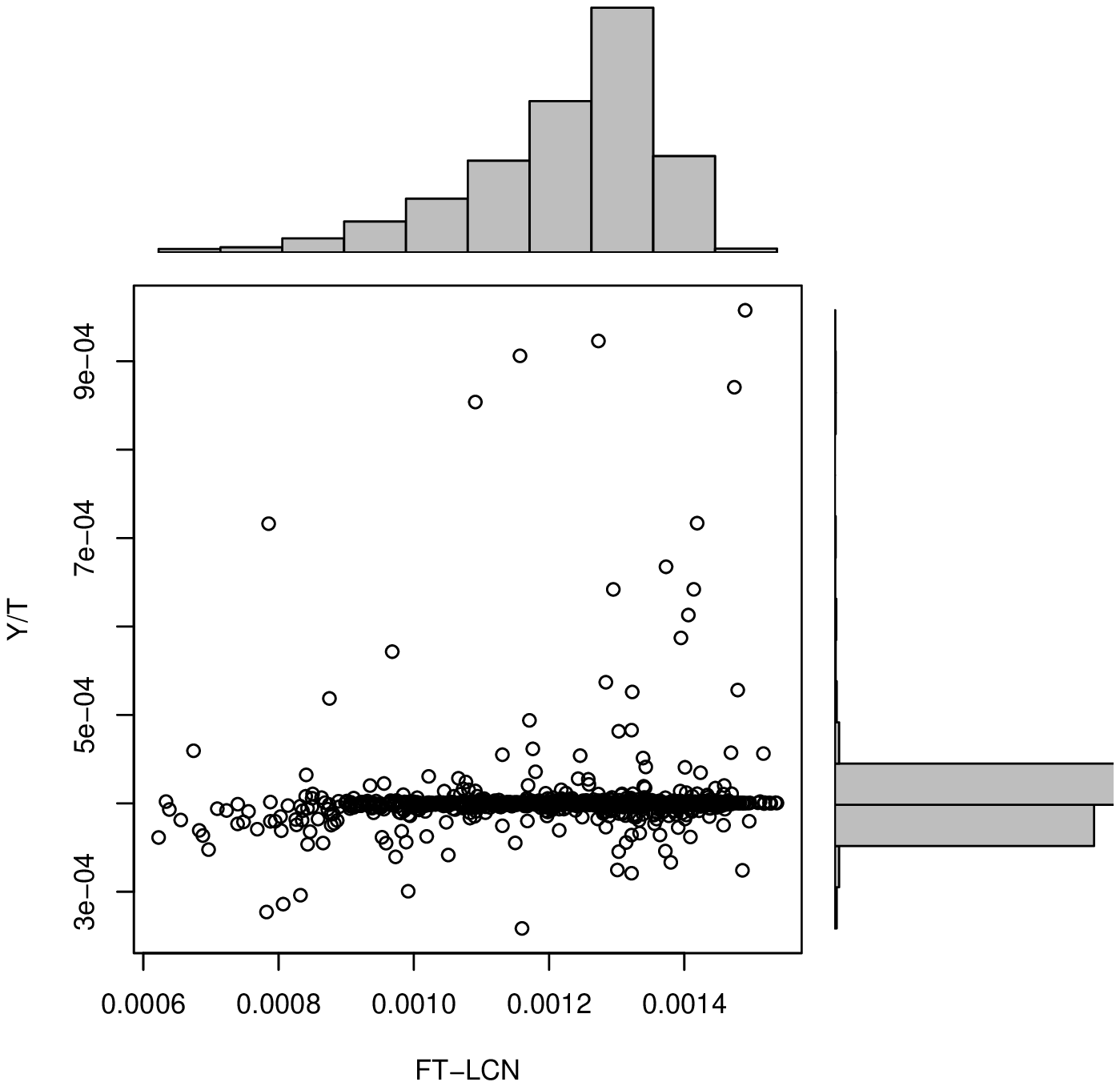}}
\end{tabular}
\caption{Correlations between $2 \overline{Y}/T$ and the largest FT--LCN for chaotic orbits (left panel) and
 between  $\overline{Y}/T$ and the largest FT--LCN for regular orbits
(right panel) for $T=5000$ u.t. The distributions of both the MEGNO and the FT--LCN are also
included. These figures were generated using \cite{Wessa09}.}
\label{figure 4.2.1}
\end{center}
\end{figure}

For the sake of making the comparison clearer, in Figs. \ref{figure 4.2.1} and \ref{figure 4.2}
we have changed the scale of $\overline{Y}$ through the division by the total integration time, $T$, in order to
have both the MEGNO and the largest FT--LCN values of the same magnitude.
Let us recall that a factor 2 should be added in the case of chaotic orbits
since we are dealing with $\overline{Y}/T$ instead of $Y/T$ (see the discussion
at the end of Section~\ref{S2}).

At short integration times we observe a few orbits, classified as chaotic by the FT--LCN, 
falling very close to the regular value of the MEGNO. Indeed, only 0.82$\%$ of the subset 
of chaotic orbits $\mathbf{Oc}$ attained MEGNO values close to $2.01$ ($\sim -3.095$ in Fig. \ref{figure 4.2.1}, left panel) at $T=5000$ u.t., 
while 8.82$\%$ of the orbits in the subset $\mathbf{Or}$
achieved MEGNO values within the range [2.01,10), indicating either their mild chaotic character or 
that the total integration time $T=5000$ u.t. is not large enough for 
the asymptotic regular value to be reached. 

For the orbits in $\mathbf{Oc}$, the mean of $\log(\mathrm{FT-LCN})\approx -1.267$ 
and the mean of $\log(2\overline{Y}/T)\approx -1.317$, while 
the corresponding standard deviations are $\approx 0.617$ and $\approx 0.676$, 
respectively, with a correlation coefficient of $r \approx 0.942$.  Thus, both
distributions are quite similar. 

For the regular sample instead, the concordance is, as expected, less fortunate. 
The concomitant correlation coefficient $r$ is close to $0.07$. 
The respective mean values are 
$\mathrm{FT-LCN}\approx 0.00126$ and $\overline{Y}/T\approx 0.0004$, with rather different 
standard deviations, namely, $0.00016$ for the distribution of the FT-LCNs and $0.000035$ 
for $\overline{Y}/T$.  

Notice must be taken of the fact that we are comparing 
values of the FT--LCNs and the MEGNO corresponding to different integration times. 
Altogether, the classification by recourse of the MEGNO provides fairly good results 
taking account that they are obtained for $T=5000$ u.t., half the total integration time 
used by \cite{MCW05} in their computation of the FT--LCN.   
The comparison of both dynamical indicators at the very same total integration time $T$ 
is the subject of the forthcoming section.

\subsection{Results at intermediate integration times}
\label{S4ab}

\begin{figure}
\begin{center}
\begin{tabular}{cc}
\hspace{-20mm}\resizebox{90mm}{!}{\includegraphics{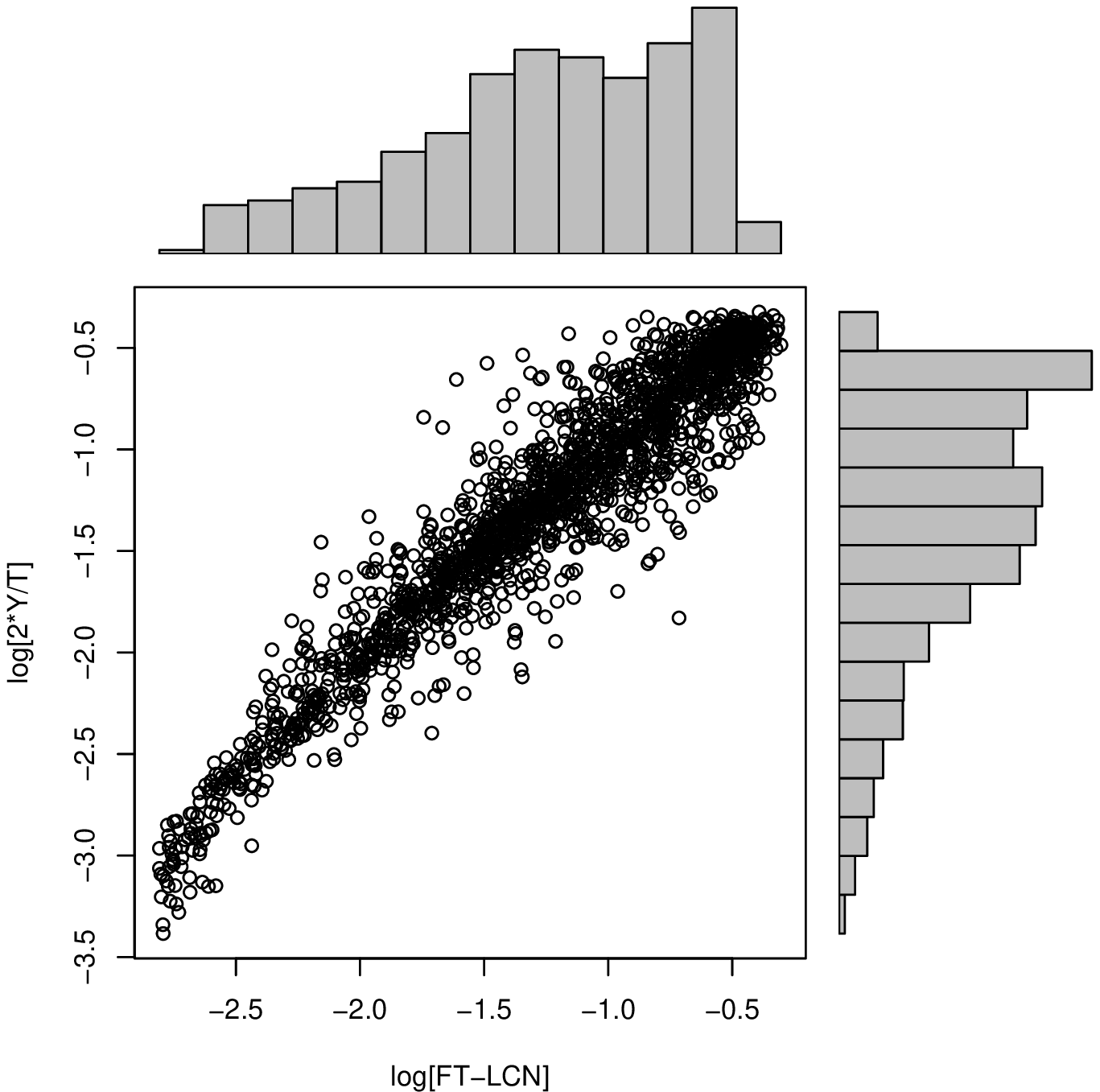}}&
\hspace{-30mm}\resizebox{90mm}{!}{\includegraphics{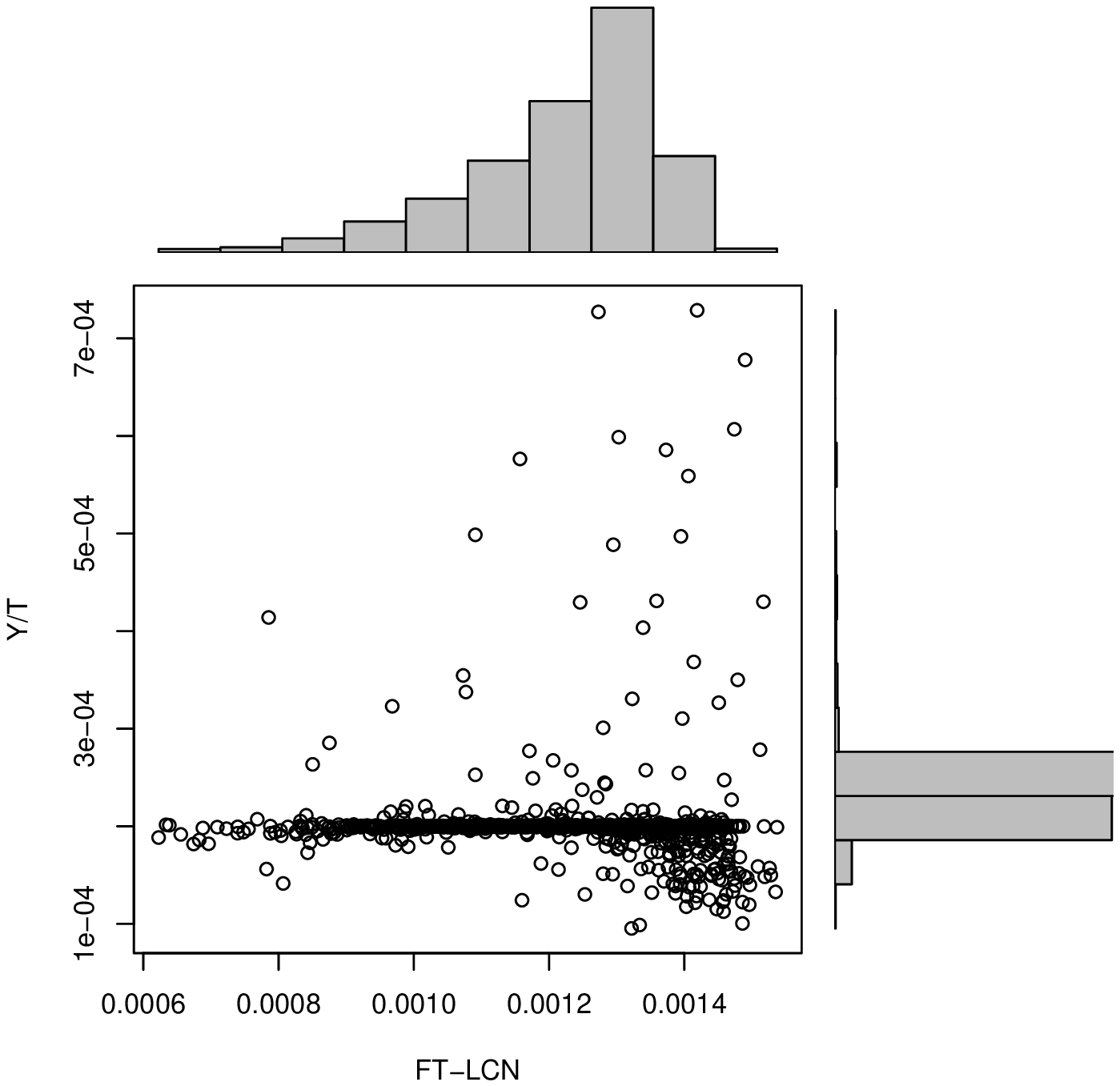}}
\end{tabular}
\caption{Correlations between $2 \overline{Y}/T$ and the largest FT--LCN
for chaotic orbits (left panel) and
 between  $\overline{Y}/T$ and the largest FT--LCN for regular orbits
(right panel) for $T=10000$ u.t. The distributions of both the MEGNO and the FT--LCN are also 
included. These figures were generated using \cite{Wessa09}.}
\label{figure 4.2}
\end{center}
\end{figure}

From Figs. \ref{figure 4.2} we observe a rather good agreement between 
the classification given by the 
MEGNO and that due to the largest FT--LCN at $T=10000$ u.t.  
In fact, the chaotic component $\mathbf{Oc}$ appointed by the FT--LCN is re-attained by means of the MEGNO, 
i.e. all  orbits in $\mathbf{Oc}$ have MEGNO values lying on the MEGNO irregularity range 
(above $\sim -3.4$ on the vertical axis in the plot on the left). 
Since we have rescaled the MEGNO by $1/T$ and $2/T$ for regular and chaotic
orbits respectively,  those values above, but close to, either $0.0002$ or $-3.4$ after 10000 u.t.,
suggest that the orbit could be proximate to an hyperbolic object (like unstable periodic orbits, 
2D hyperbolic torus). On the other hand, for MEGNO values $\lesssim 0.0002$, the orbit may be close 
to elliptical objects (stable periodic orbits, 2D resonant elliptical torus),  
as it is shown in \cite{CGS03}. 

On the left panel in Fig. \ref{figure 4.2}, corresponding to chaotic orbits,  
we distinguish a clearly linear correlation between 
$\log(2 \overline{Y}/T)$ and $\log(\mathrm{FT-LCN})$ for $T=10000$ u.t. Indeed, the   
correlation coefficient is $r\approx 0.95$; the mean value of $\log(2\overline{Y}/T)\approx -1.29$,
while the concomitant mean value of $\log(\mathrm{FT-LCN})\approx -1.27$.

In the same figure on the right, we show the correlation between $\overline{Y}/T$ and the largest FT--LCN 
for the orbits in $\mathbf{Or}$.  
In this case, the correlation coefficient is $r\approx -0.026$, 
the mean value of $\overline{Y}/T$ is rather close to $0.0002$ with a standard deviation
$\lesssim 10^{-5}$, while for the FT-LCNs the mean value is $0.0013$ with
a standard deviation of order $\approx 10^{-4}$. 
Let us point out the sharp character of the distribution of the MEGNO values around the predicted one 
for regular motion, while the FT--LCNs' distribution is rather blunt, as the standard deviation of 
both distributions indicate. This fact should be deemed as an 
advantageous feature of the MEGNO over the largest FT--LCN. 

Notice must be taken that 
the empirical value $V_c^e=0.00155$ adopted by \cite{MCW05} for their classification, 
is greater than the mean value of the largest FT--LCN for regular orbits,  
which indeed might be a misleading factor for the task.

Thus, from the regular component $\mathbf{Or}$, which encompasses 1644 orbits on the whole, 
1513 orbits have values of the $\overline{Y}/T$ in the interval $[0.0001;0.000201)$, 
in due accord with their stable, regular character.  
However, a discrepancy is found for a subset $\mathbf{Or}_d$, including the remaining 131 orbits 
classified as regular by their largest FT--LCN, whose $\overline{Y}/T$ values,  
however, lie within the range $[0.000201;0.001)$ revealing their possible irregular character. 

\begin{figure}
  \begin{center}
    \begin{tabular}{cc}
    \hspace{-5mm}\resizebox{63mm}{!}{\includegraphics{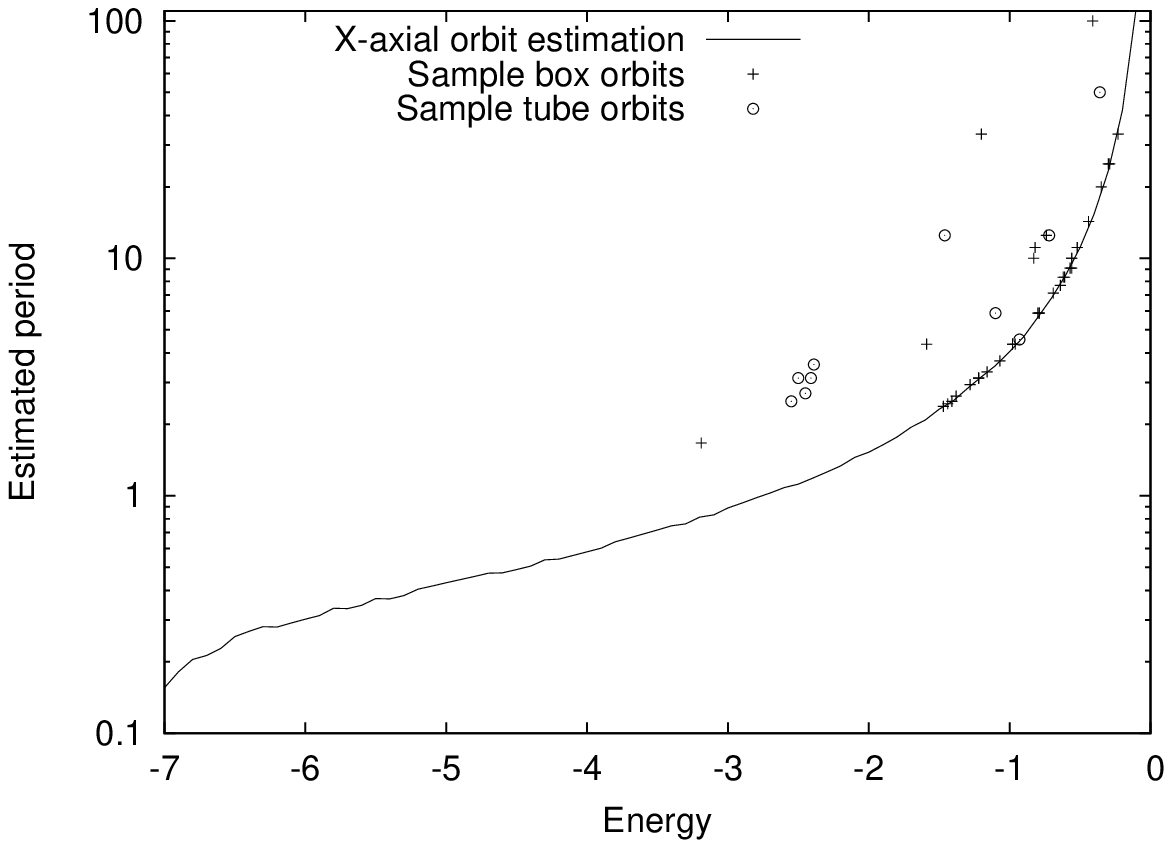}}& 
    \hspace{-5mm}\resizebox{63mm}{!}{\includegraphics{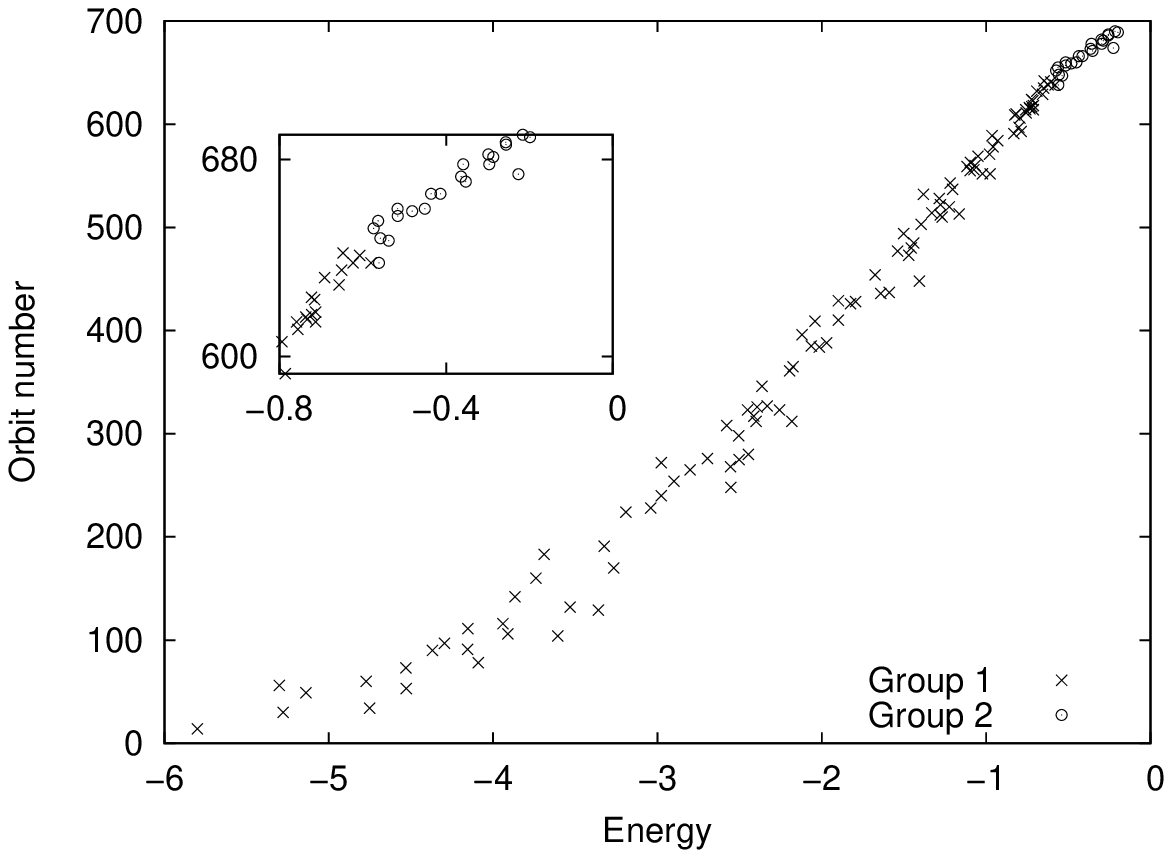}}    
    \end{tabular}
   \caption{Period of the x--axis orbit as a function of the energy adopted as 
    $T_c(E)$ for box orbits, along with the approximate period of a sample of both box 
    and tube orbits (on the left). 
    Energy of the 131 orbits in $\mathbf{Or}_d$ (on the right).}
    \label{figure 4.3}
  \end{center}
\end{figure}

An issue to be stressed is the fact that the deemed orbits belong to different energy surfaces and, 
for each energy, a different characteristic time--scale, $T_c(E)$, can be defined. 
Moreover, in order to ensure that both the FT--LCN and MEGNO are well computed for a given orbit, 
the total integration time should verify $T>>T_c(E)$.
Therefrom, on fixing the condition $T\geq10^3T_c(E)$ to obtain
confident values for both indicators, we conclude that only those orbits with $T_c(E)\leq 10$ 
would be properly classified for a total integration time of 10000 u.t. 
The function  $T_c(E)$ for box orbits, taken as the period of the stable x--axis periodic orbit, 
is plotted in Fig. \ref{figure 4.3} on the left, which shows that  $T_c(E)= 10$ 
corresponds to an energy value $E\sim -0.58$.
We have computed the approximate period of some box orbits (crosses in the figure) in 
order to test the suitability of the adopted time--scale  $T_c(E)$ for boxes. 
The approximate period of some tube orbits are also included in the figure.

The energy values corresponding to the orbits in $\mathbf{Or}_d$ 
are displayed on the right of Fig. \ref{figure 4.3}, 
where we have labeled as group 1 those with 
energies $E<-0.58$, and as group 2 the ones for which $E >-0.58$.

From the 131 orbits in $\mathbf{Or}_d$ 41 are tubes 
while 90 are boxes, 20 of which have $E >-0.58$, i.e. their dynamical indicators would 
still be in a transient phase. 
On the whole, we count 70 box orbits in $\mathbf{Or}_d$ satisfying the condition 
$T>>T_c(E)$.  

Let us remark that all orbits in $\mathbf{Or}_d$ have $2\lesssim\overline{Y}\lesssim 7$ at
$T=10000$ u.t. so, even when they could evince some local instability, 
they behave as stable orbits from a physical point of view. 
Nonetheless, since our aim is to subject the MEGNO to a rigorous test as 
a dynamical indicator, we will study this subset of orbits in particular.

Therefore, for the $131$ orbits in $\mathbf{Or}_d$ 
we recalculate the MEGNO but for 100000 u.t. 
to find that $52$ orbits, having MEGNO values smaller than $3.5$ at $T=10000$, approach the regular 
value $2$ at $T=100000$, while the remaining $79$  attain greater values of the MEGNO for the 
larger integration time. Let us mention that for some orbits the MEGNO value is barely higher
than the adopted threshold of $2.01$ (as it will be shown in section \ref{S4abcd}).
In the following subsection, we will identify the first group by $\mathbf{Or}_d^s$ and the second one by $\mathbf{Or}_d^u$.

In order to determine the actual character of the orbits in $\mathbf{Or}_d$,
we will recourse to a slight variation of the so--called Fast Lyapunov Indicator (FLI) 
(see \cite{FGL97}).
Briefly, the FLI is defined as the supremum of the norm of the tangent vector $\vec{\delta}$.
Thus, we will follow the evolution of $\langle\delta(t)\rangle$ where

$$\langle\delta(t)\rangle=\frac{1}{t}\int_0^t \delta(t)dt,$$

\noindent
with $\delta = \|\vec{\delta}\|$ and 
$\vec{\delta}$ is the solution of the variational equations, taking as initial value $\delta_0=10^{-4}$.

\begin{figure}[t!]
\begin{center}
\begin{tabular}{cc}
\hspace{-6mm}\resizebox{65mm}{!}{\includegraphics{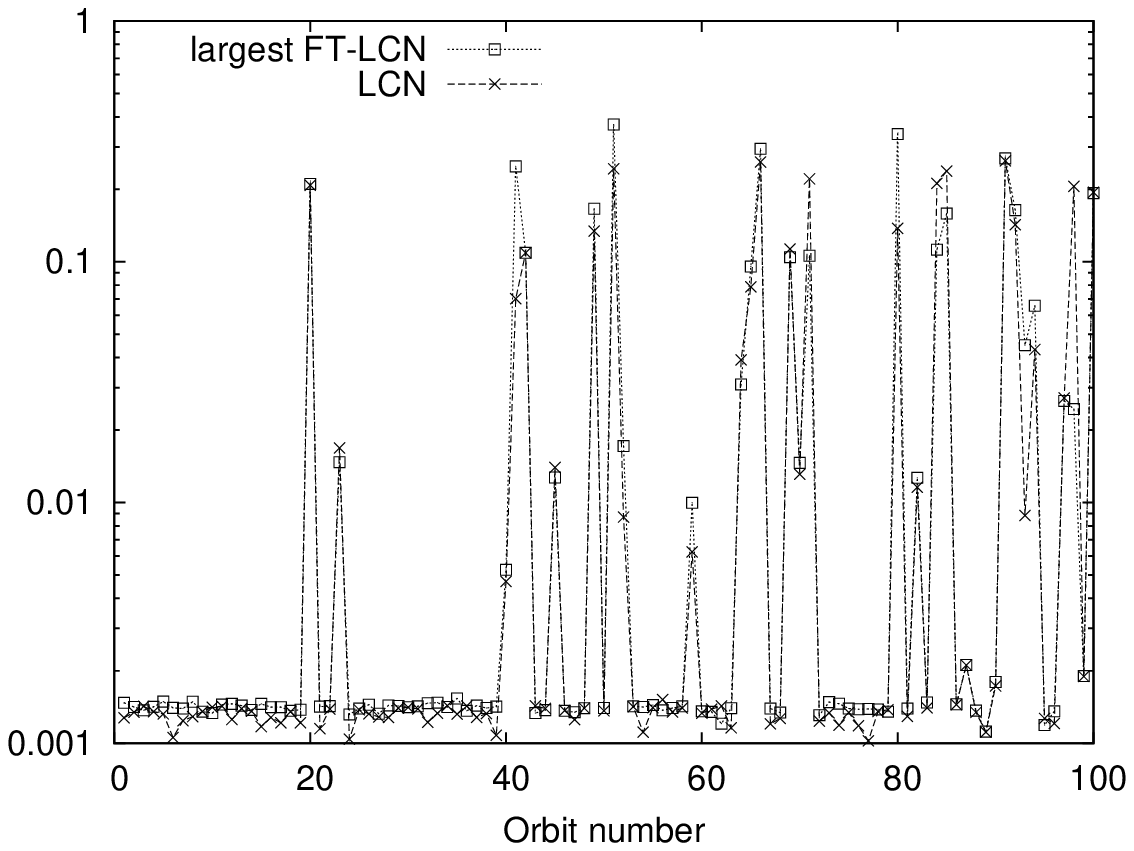}}&
\hspace{-6mm}\resizebox{65mm}{!}{\includegraphics{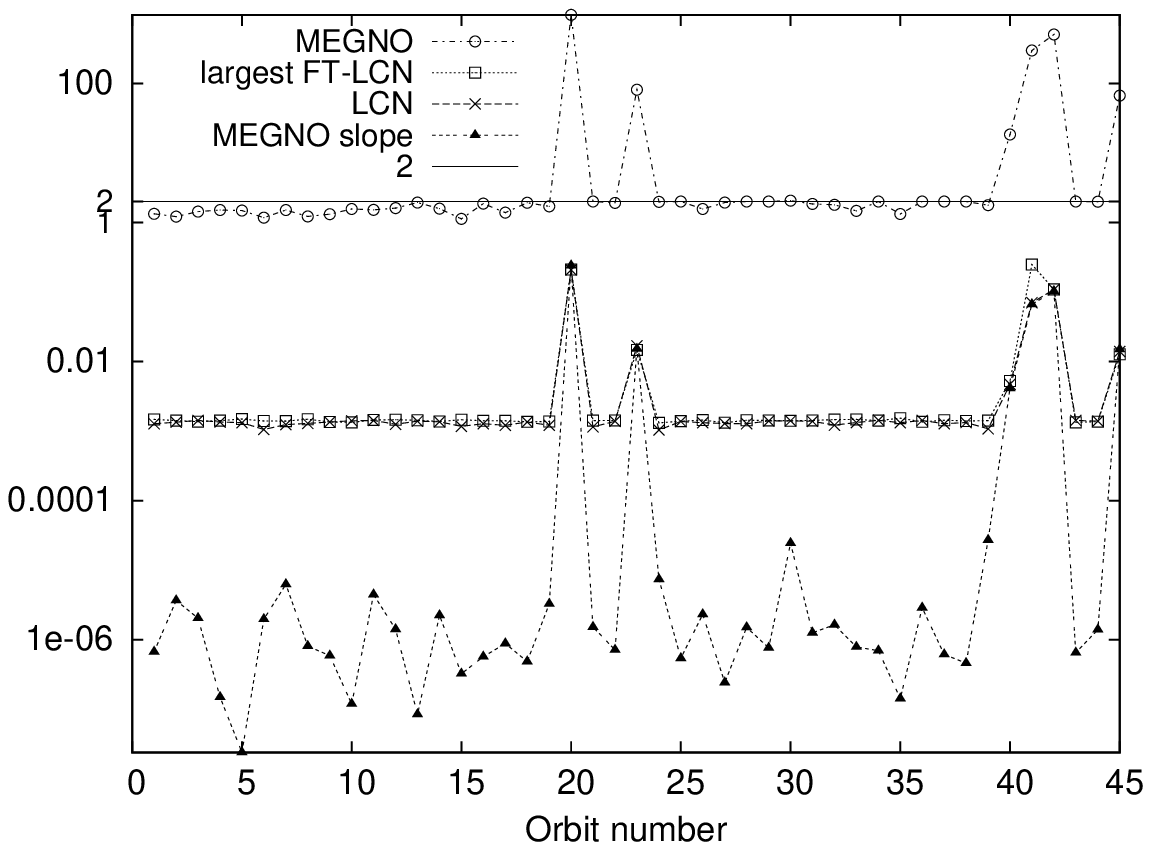}}
\end{tabular}
\caption{Both the largest FT--LCN provided by \cite{MCW05} and the maximum LCN 
 computed by our using the classical algorithm for a sample of 100 orbits in $\mathbf{O}$~(left). 
 Largest FT--LCN, LCN, slope of the MEGNO and MEGNO for a smaller sample of 45 orbits~(right).}
\label{fig4.5}
\end{center}
\end{figure}

Since motion times larger than 10000 u.t. will be considered in the next section, let us first compute the LCN for a sample of 
100 orbits in $\mathbf{O}$ for $T=10000$ u.t. using the classical algorithm of \cite{BGS76}, and compare the obtained values  with the largest FT--LCN derived by \cite{MCW05}.
The result of the comparison is illustrated in the plot on the left of  Fig.~\ref{fig4.5}, which gives account 
of an actually quite good agreement. 
This encourages us to compute the LCNs and assimilate them with the largest FT--LCNs for $T=100000$ u.t.
In Fig.~\ref{fig4.5} on the right we also include the estimation of the largest FT--LCN derived from the slope 
of the MEGNO and the MEGNO itself for a small sample of orbits (45 on the whole). 
Let us point out that the slope of the MEGNO yields a better estimation of the largest FT--LCN, 
particularly for the regular orbits, for which it lies below $10^{-4}$, which is much smaller 
than both $V_c^e$ and $V_c^t$ for $T=10000$ u.t. 
Nonetheless, we will restrict our comparative analysis to the MEGNO, the LCN and the mean FLI 
($\langle\delta(t)\rangle$), since deriving the expected theoretical value for 
the slope of the MEGNO in case of quasi-periodic motion is difficult. 
In fact, the same occurs with the mean FLI, for which it is not possible to determine an asymptotic value 
whenever the orbit is confined to a torus.
Indeed, for such a quasi--periodic orbit, $\gamma_{q}$, the solution of the variational equation in 
$\mathbb{R}^{2N}$ can be recast as 
$$\delta\left(\gamma_{q}(t)\right)\approx\delta_0\left[1+w_{q}(t)+
t\left(\lambda_{q}+u_{q}(t)\right)\right],$$
where $\lambda_{q}>0$ is the linear rate of divergence around $\gamma_{q}$, and
$w_{q}(t)$ and $u_{q}(t)$ are oscillating functions of $t$ of bounded amplitude 
(in general quasi--periodic and with zero average), 
satisfying $|u_{q}(t)|\le b_{q}<\lambda_{q}$. 
The parameter $\lambda_q$ is a measure of the lack of isochronicity around the orbit 
since it is related to the maximum eigenvalue of the matrix $\partial\mathbf{\omega}/\partial{\mathbf{I}}$, 
$\mathbf{\omega}$ and $\mathbf {I}$ being the frequency and action vectors associated to the torus,
respectively (for an isochronous system, such as the harmonic oscillator,
$\lambda=b_{q}=0$ for all $\gamma$).


\subsection{Results at large integration times}
\label{S4abc}

\begin{figure}[ht!]
\begin{center}
\begin{tabular}{cc}
\hspace{-5mm}\resizebox{63mm}{!}{\includegraphics{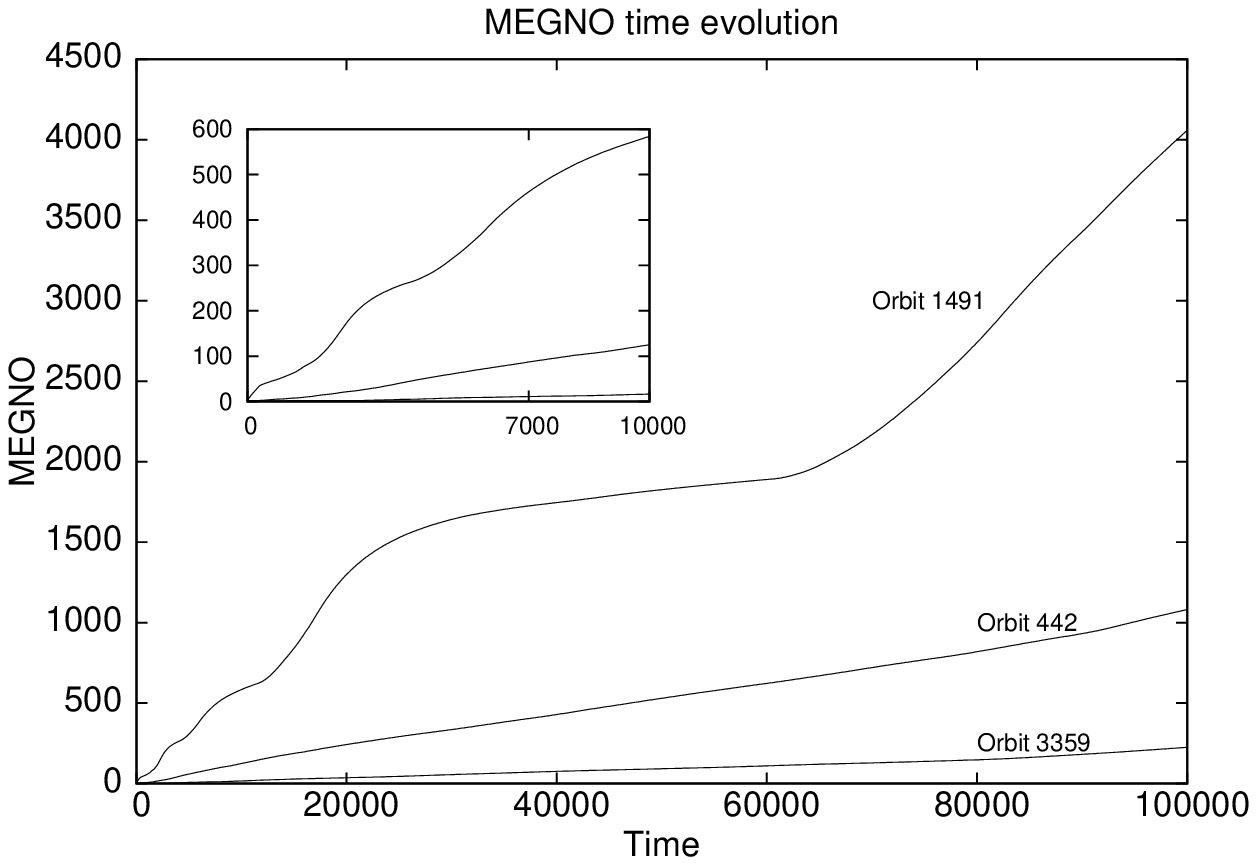}}&
\hspace{-5mm}\resizebox{63mm}{!}{\includegraphics{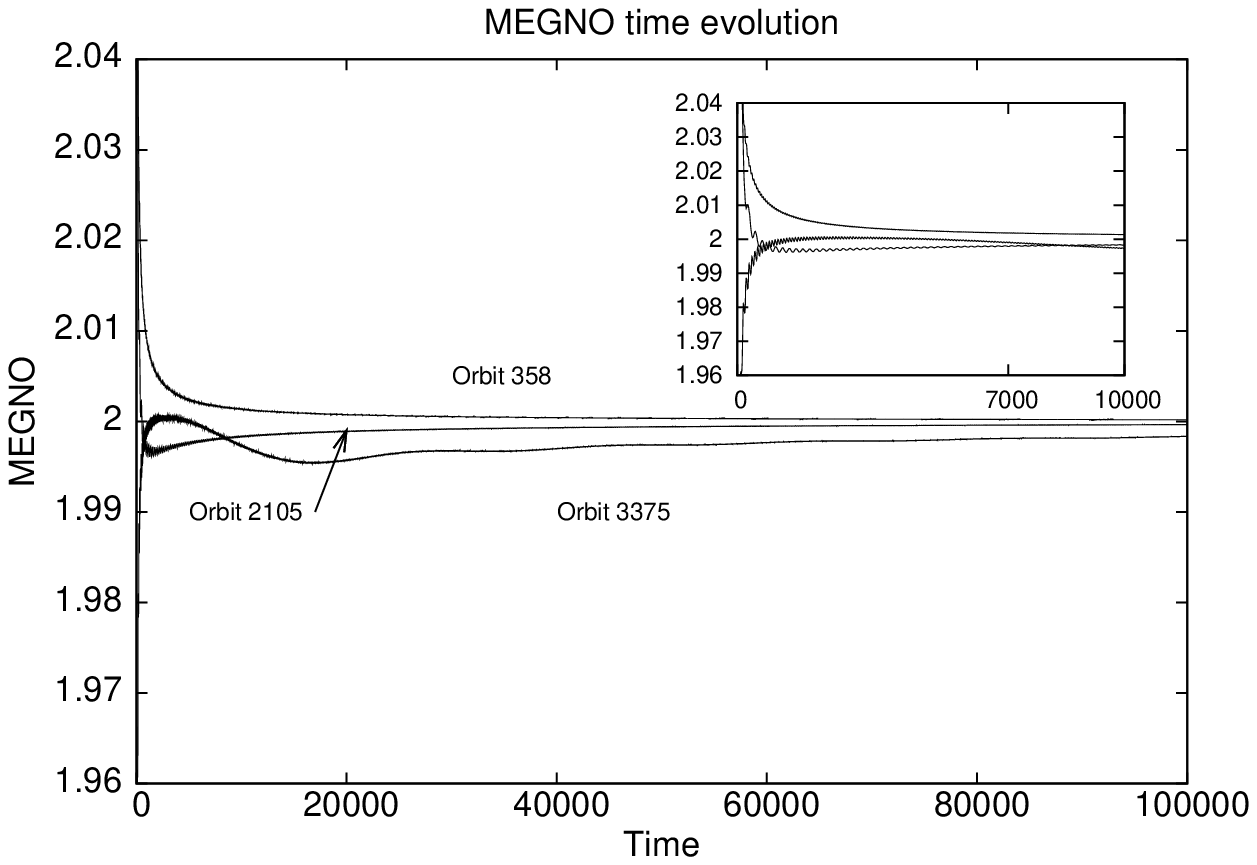}}
\end{tabular}
\begin{tabular}{cc}
\hspace{-5mm}\resizebox{63mm}{!}{\includegraphics{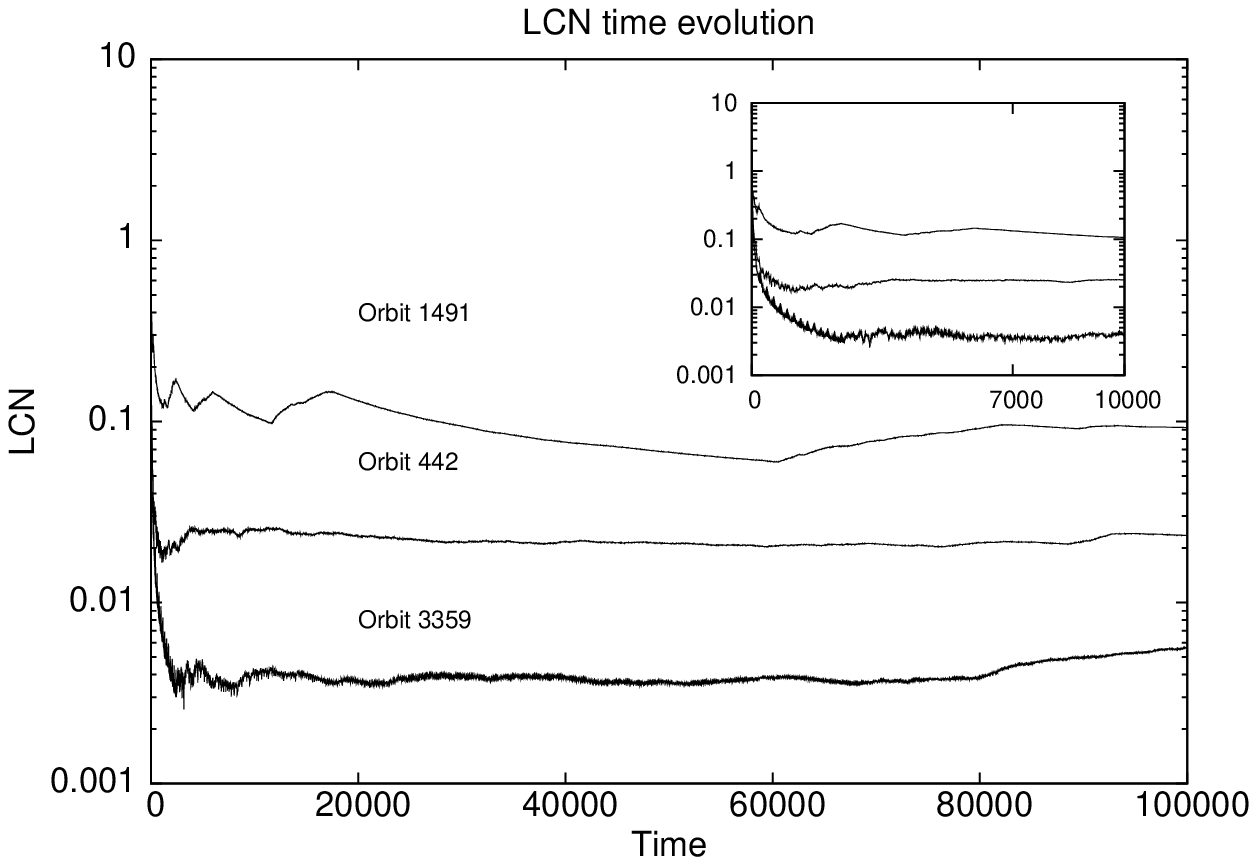}}&
\hspace{-5mm}\resizebox{63mm}{!}{\includegraphics{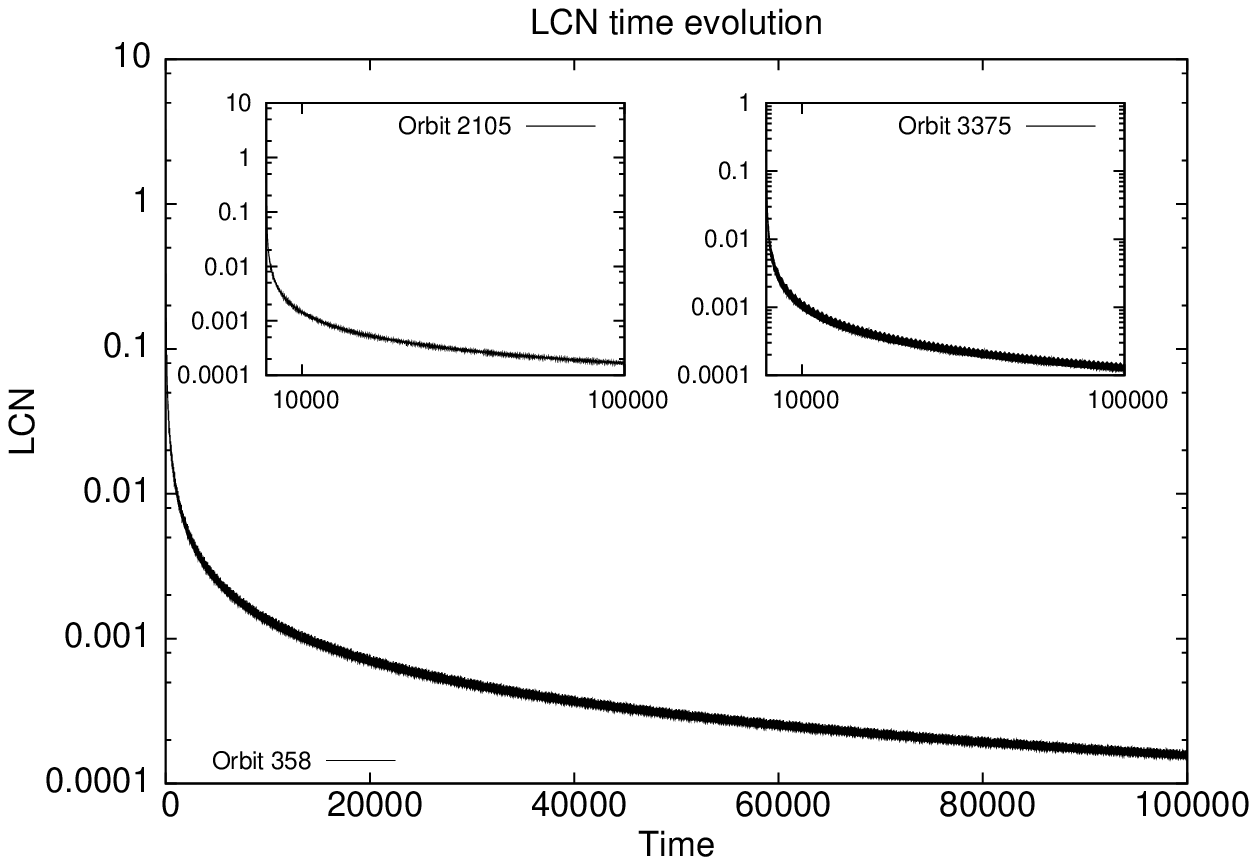}}
\end{tabular}
\begin{tabular}{cc}
\hspace{-5mm}\resizebox{63mm}{!}{\includegraphics{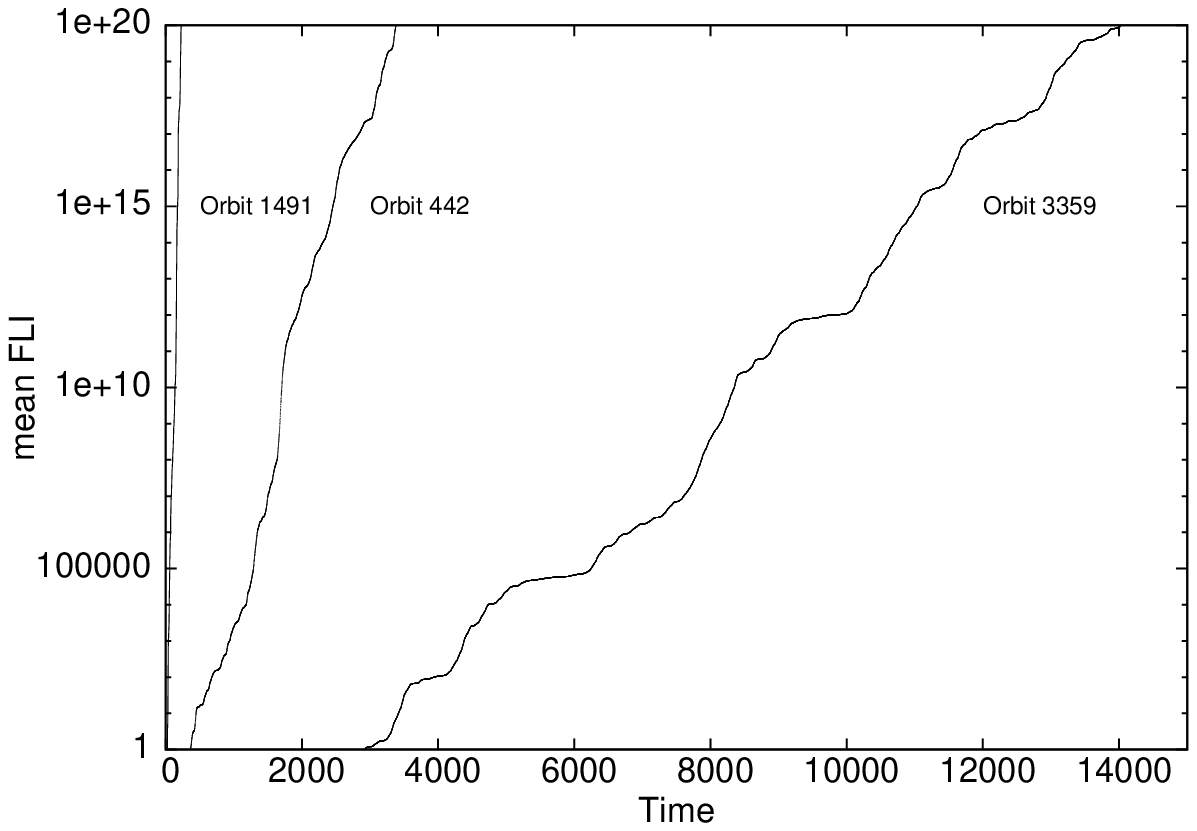}}&
\hspace{-5mm}\resizebox{63mm}{!}{\includegraphics{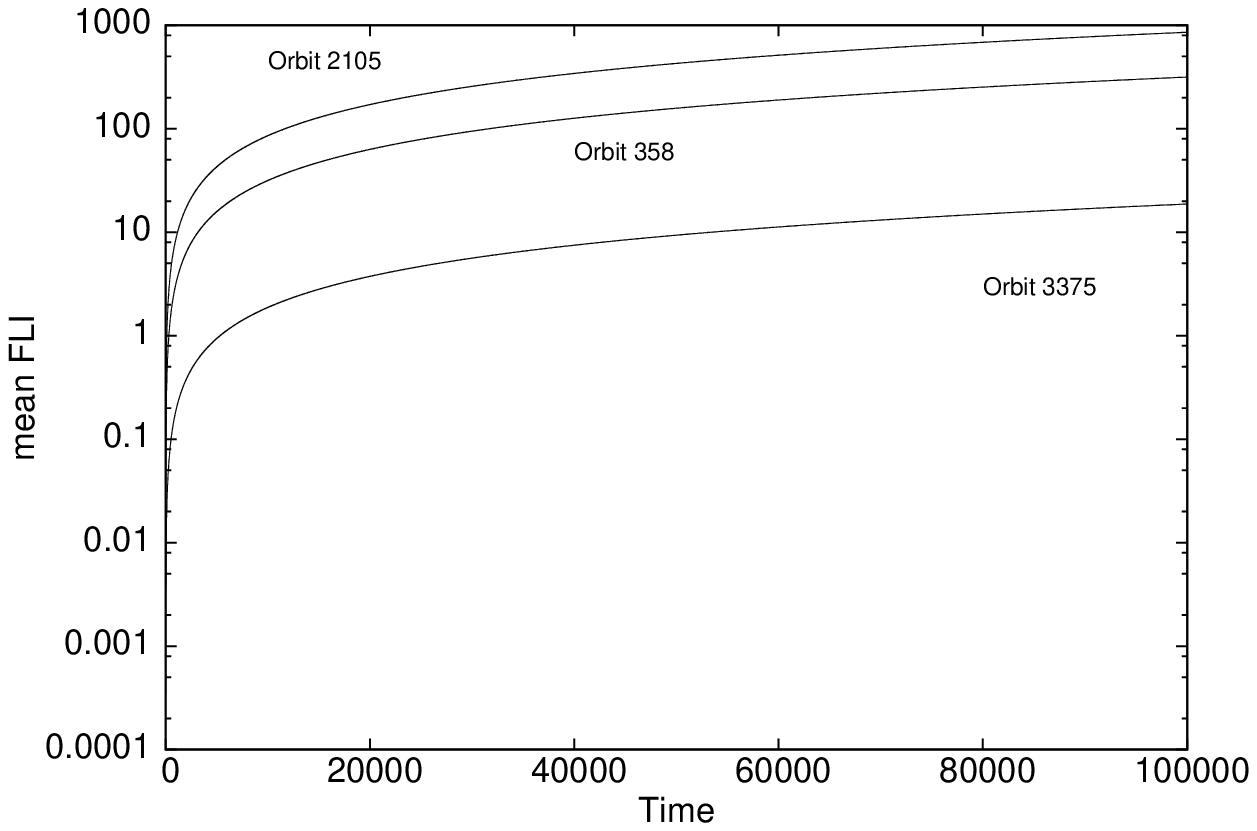}}
\end{tabular}
\caption{Typical temporal evolution of the MEGNO, the LCN and
the mean FLI for chaotic and regular orbits (e.g. in
$\mathbf{Oc}$ and $\mathbf{Or}-\mathbf{Or}_d$, respectively). 
For the plots in the last panel, 
the runs were stopped at $\langle\delta(t)\rangle=10^{20}$ in order to avoid overflow.}
\label{fig4.54}
\end{center}
\end{figure}

In the present section we will be concerned with the temporal evolution of the three indicators to be 
compared, namely the MEGNO, the LCN and the mean FLI (for which no re-normalization was performed   
and in the case of the exponential growth of $\delta(t)$  
the integration was stopped at $\langle\delta(t)\rangle=10^{20}$), 
for large motion times. 
Figs.~\ref{fig4.54} displays the typical behavior of these indicators
for chaotic and regular orbits which are identified by their orbit number.  
For the illustration the orbits 1491, 442 and 3359 from the set $\mathbf{Oc}$   
and  358, 2105 and 3375 from $\mathbf{Or}-\mathbf{Or}_d$ 
have been selected.
 
The MEGNO shows a linear growth with time for the orbits in $\mathbf{Oc}$, 
except for orbit 1491 for which exhibits two similar linear trends and a flat behavior 
between $\sim 20000$ and $\sim 60000$ u.t., suggesting that during this time interval
the orbit may be close to some elliptic structure. 
On the other hand, for the orbits in $\mathbf{Or}-\mathbf{Or}_d$ 
the MEGNO asymptotically approaches the predicted value, 2, both at 10000 u.t and 100000 u.t. 

In regards to the LCN, a similar behavior is observed for the orbit 1491, while for 
the regular orbits converges  
to the theoretical expected value $V_c^t=\ln T/T\sim 1.2\times 10^{-4}$. 

Finally, for those orbits in $\mathbf{Oc}$ the mean FLI  displays an almost exponential dependence with time 
(in fact it goes as  $e^{\sigma t}/t$), while it attains much lower values for orbits in 
$\mathbf{Or}-\mathbf{Or}_d$ for which it depends with time in a linear fashion   
 (note the logarithmic scale in the vertical axis),   

Thus, these figures provide information about the expected  behavior of the three indicators 
in the cases of both regular and chaotic motion, which will be of use 
 to determine the character of those orbits in $\mathbf{Or}_d^s$ and $\mathbf{Or}_d^u$. 
Let us recall that we will restrict our study to those orbits satisfying  the condition $E\le -0.58$ 
and for which a good estimation of their period is at hand.

\begin{figure}[ht!]
\begin{center}
\begin{tabular}{cc}
\hspace{-5mm}\resizebox{63mm}{!}{\includegraphics{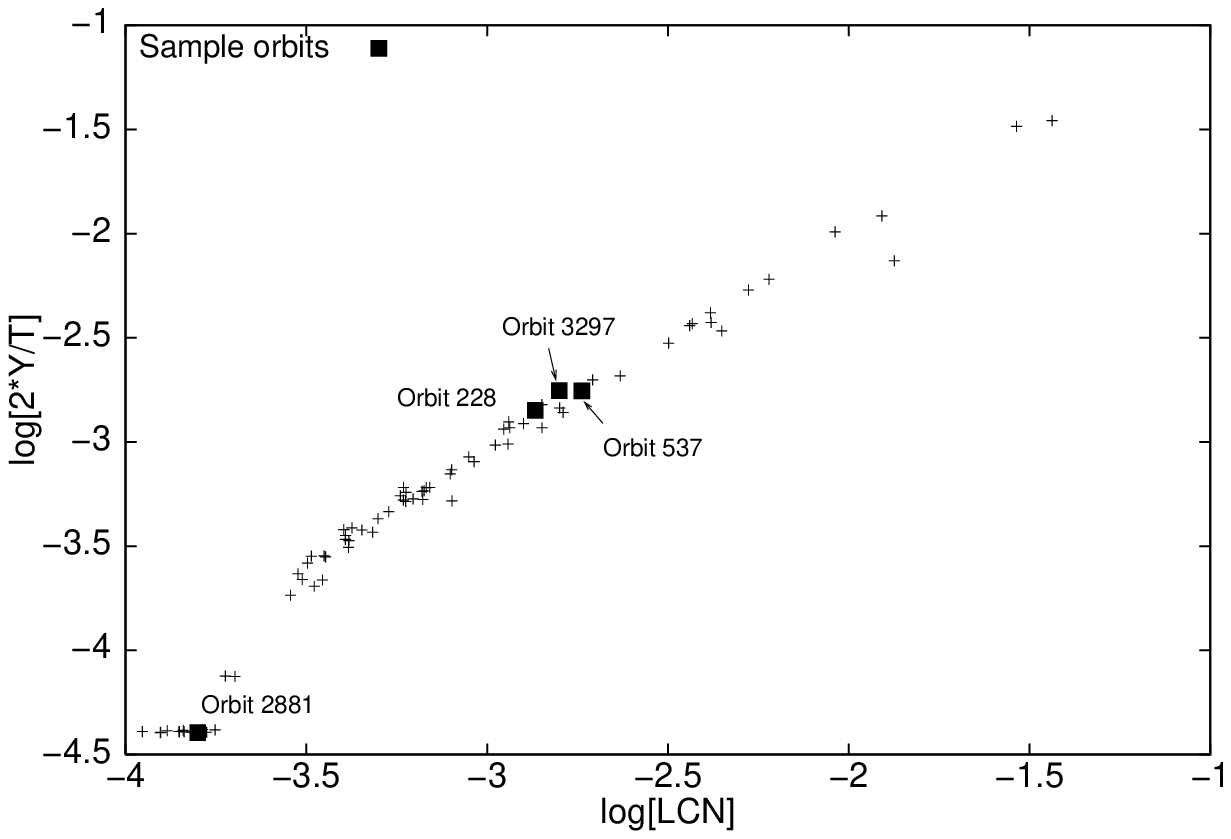}}& 
\hspace{-5mm}\resizebox{63mm}{!}{\includegraphics{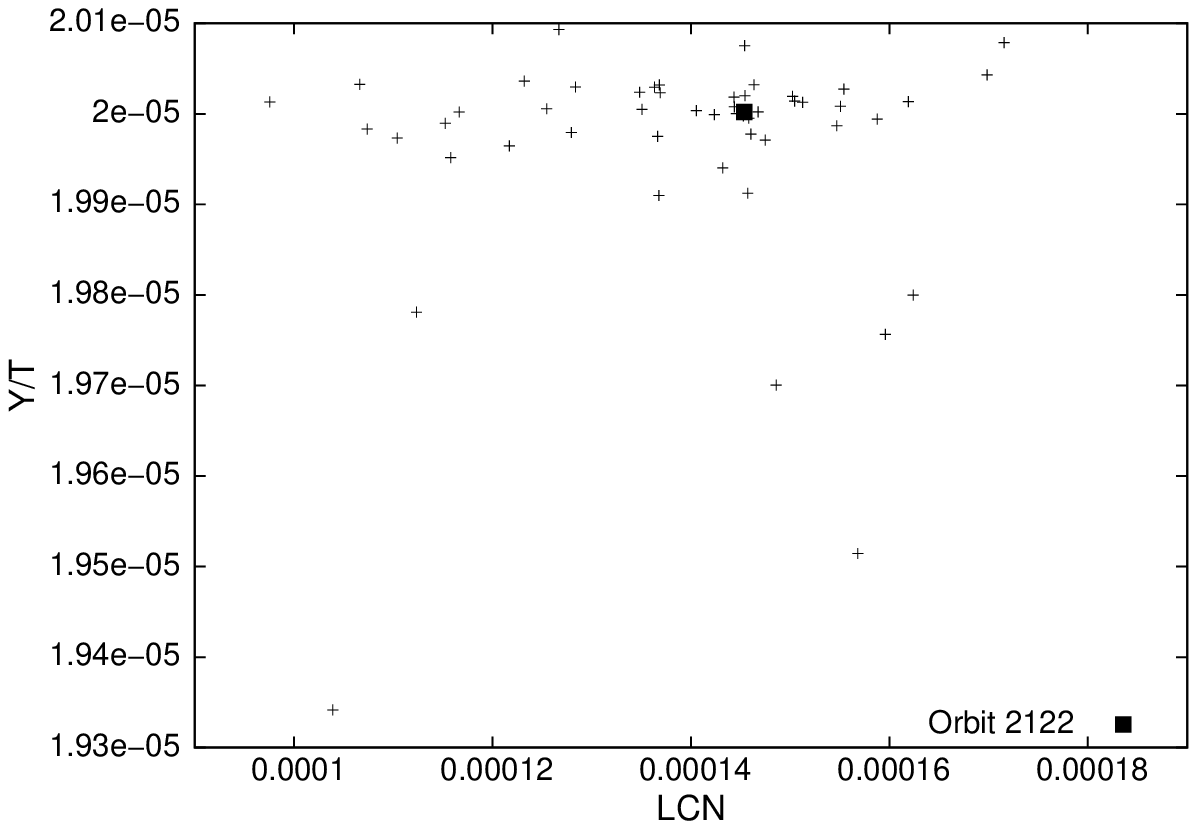}}
\end{tabular}
\caption{Correlations between $2 \overline{Y}/T$ and the LCN
for orbits in $\mathbf{Or}_d^u$ (left panel) and
between $\overline{Y}/T$ and the LCN for orbits in $\mathbf{Or}_d^s$
(right panel) at $T=100000$ u.t. The orbits marked with a full dot will be studied separately, following the temporal evolution of the three dynamical indicators.}
\label{figure 4.5}
\end{center}
\end{figure}

The correlation between the values of MEGNO and the LCN at $T=100000$ u.t.
for orbits in $\mathbf{Or}_d^u$ and $\mathbf{Or}_d^s$ is presented in Fig.~\ref{figure 4.5}.  
For the orbits in $\mathbf{Or}_d^u$, the mean value of $\log(2\overline{Y}/T)\sim -3.28$ while the
mean of $\log(\mathrm{LCN})\sim -3.13$, and the standard deviations are 0.72 and 0.56 respectively, 
 the correlation coefficient being close to 0.98. 
Therefore, not only the correlation between both indicators is quite good, but the two first moments 
of their concomitant distributions are rather similar as well. 
Let us mention that, though we are 
computing the MEGNO for very large times, most of the orbits in $\mathbf{Or}_d^u$ attain values in the range 
$-4.5\lesssim\log(2\overline{Y}/T)\lesssim -2.5$, revealing that these orbits are mild chaotic, for which the 
mean Lyapunov time is $T_{Lyap}\sim 1500$ u.t., their characteristic period being smaller than $10$ u.t.

Meanwhile, for orbits in $\mathbf{Or}_d^s$, the mean value
of $\overline{Y}/T\approx 2\times 10^{-5}$ with a standard deviation of $1.3\times 10^{-7}$, while the mean
LCN is close to $0.00014$ with a standard deviation of about $1.7\times 10^{-5}$.  The corresponding
correlation coefficient is $0.1$. Again we point out the sharp distribution of $\overline{Y}/T$ around the
expected theoretical value. The standard deviation of both distributions 
differ in two orders of magnitude 
(notice should be taken of the different scales onto the vertical and horizontal
axis in the right plot of Fig.~\ref{figure 4.5}). 
At $T=100000$ u.t. we attain values of  
$\overline{Y}/T$ that provides a fairly good estimation of the true LCN, $\sigma=0$,
namely, of order $10^{-5}$, while the LCN computed by recourse to the standard algorithm barely abuts $10^{-4}$.

The explanation is clear; the factor $\ln T\approx 11$ in  $V_c^t$ is the responsible for this
slower convergence of  $\sigma_1$ to $\sigma=0$ as
$T\to\infty$, and $\overline{Y}/T$ tends to $\sigma$ faster than $\ln T/T$.
Indeed, for stable motion and integration times of order of,
or larger than $T\approx 22000$ u.t., $|\overline{Y}/T|\lesssim 10^{-4}$, while the LCN computed
using the standard algorithm yields $|\sigma_1|\lesssim 10^{-3}$. 
In fact, 
$${{\overline{Y}/T}\over\sigma_1}\approx {2\over \ln T}\to 0, \qquad T\to\infty.$$

Finally, the full dots in Fig.~\ref{figure 4.5} 
 correspond to five orbits selected as samples of $\mathbf{Or}_d^s$ and $\mathbf{Or}_d^u$,  
for which the study of the temporal evolution
of the three indicators for $T=100000$ u.t. will serve to determine their dynamical behavior. 
This issue will be undertaken in the forthcoming subsection.

\subsection{On the $\mathbf{Or}_d$ orbits}
\label{S4abcd}

Let us  be concerned with the detailed study of  some sample orbits of $\mathbf{Or}_d$. 
First we will aim our attention at orbits 2881, 537 and 3297 belonging to $\mathbf{Or}_d^u$, and
  2122 $\in\mathbf{Or}_d^s$. The study of orbit 228  $\in\mathbf{Or}_d^u$ will be
addressed separately.
\begin{figure}[ht!]
\begin{center}
\begin{tabular}{cc}
\hspace{-5mm}\resizebox{63mm}{!}{\includegraphics{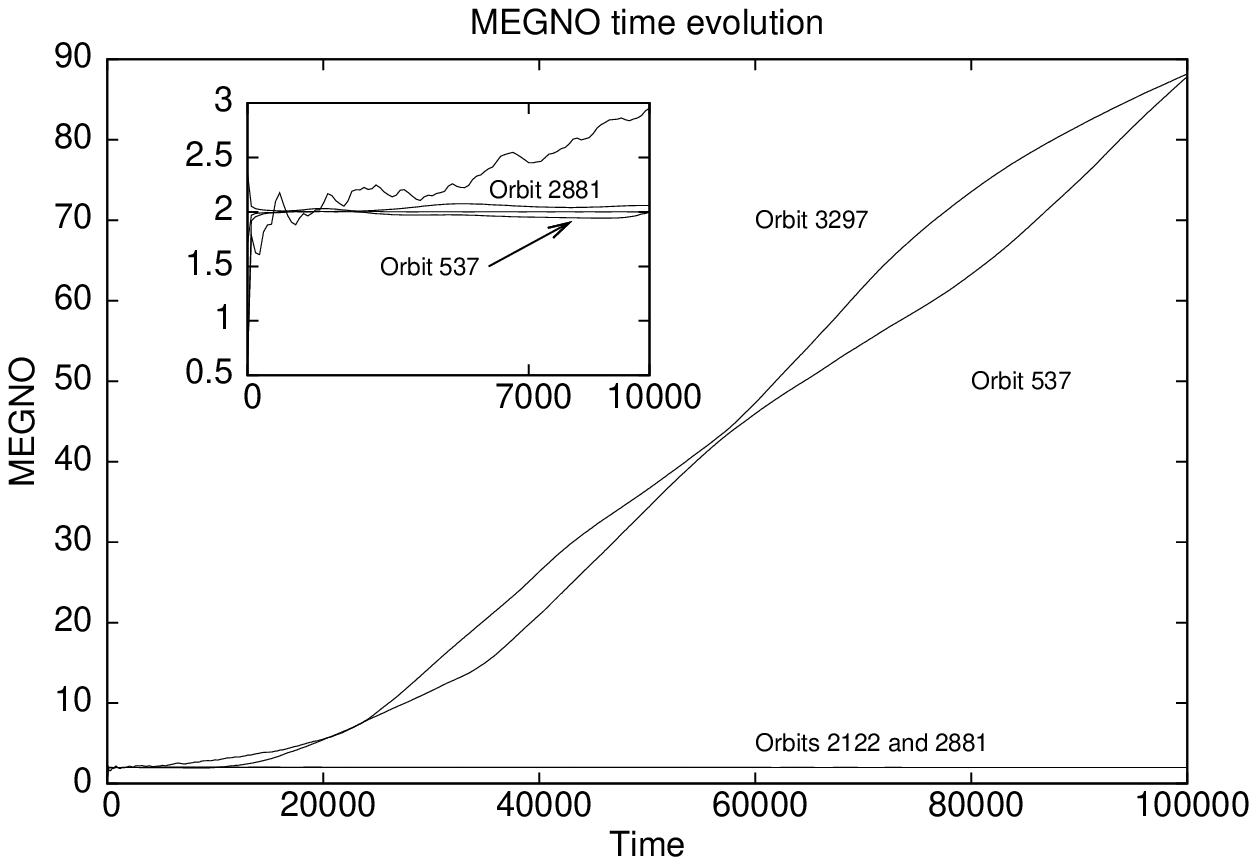}}&
\hspace{-5mm}\resizebox{63mm}{!}{\includegraphics{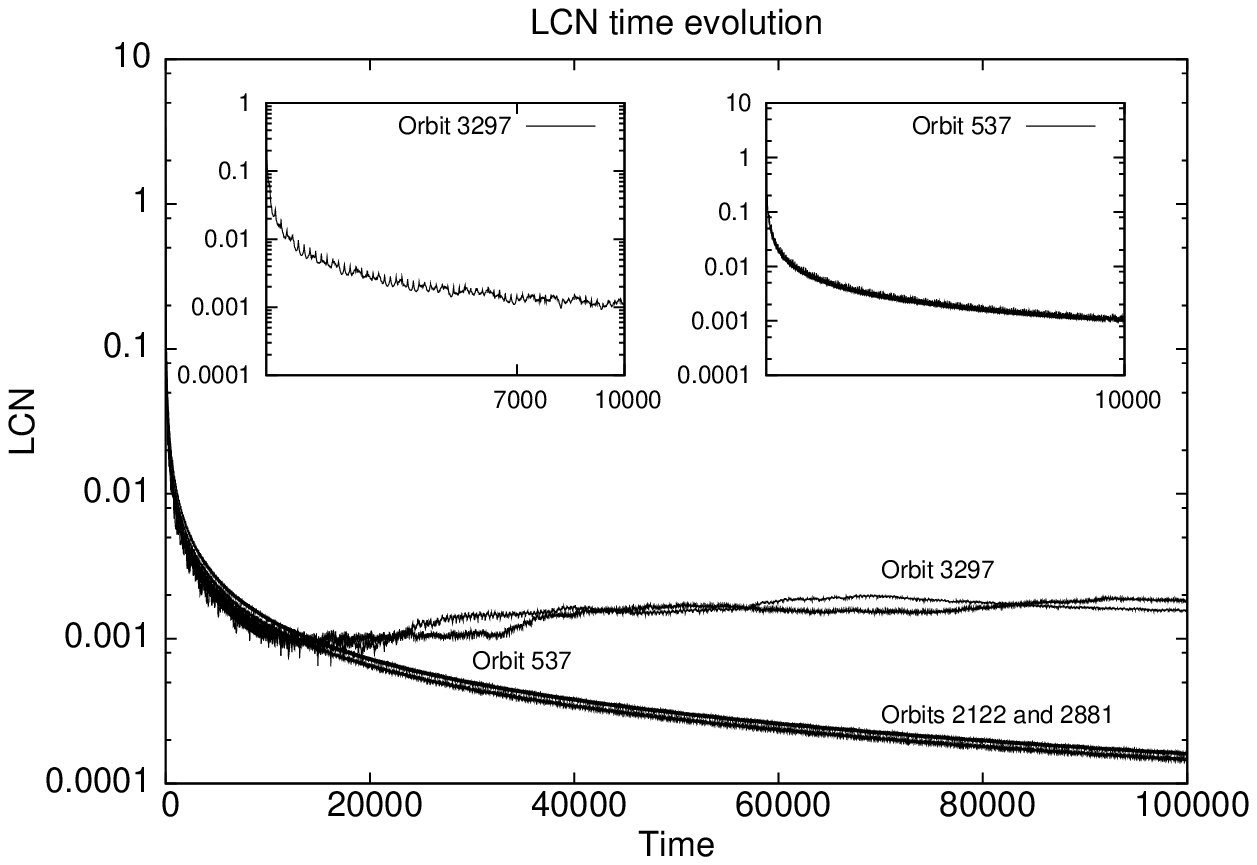}}
\end{tabular}
\begin{tabular}{cc}
\hspace{-5mm}\resizebox{63mm}{!}{\includegraphics{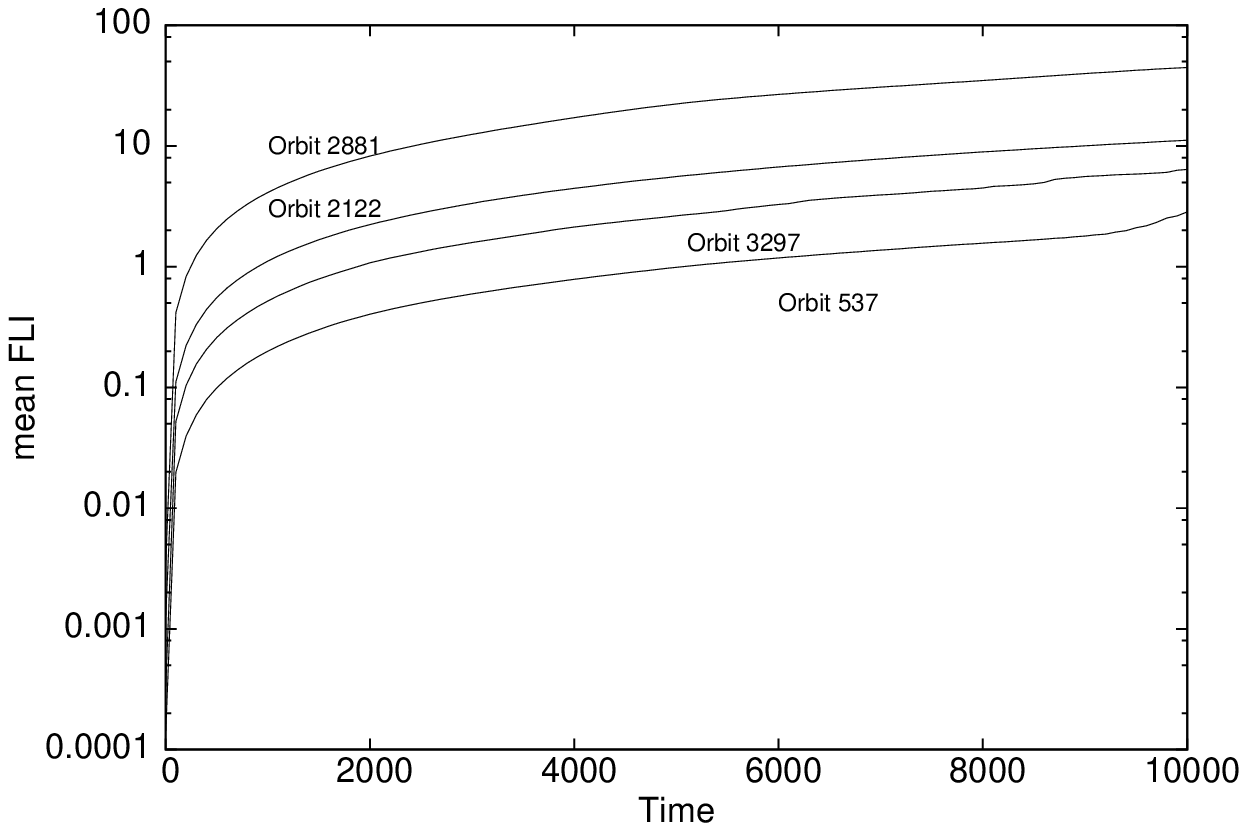}}&
\hspace{-5mm}\resizebox{63mm}{!}{\includegraphics{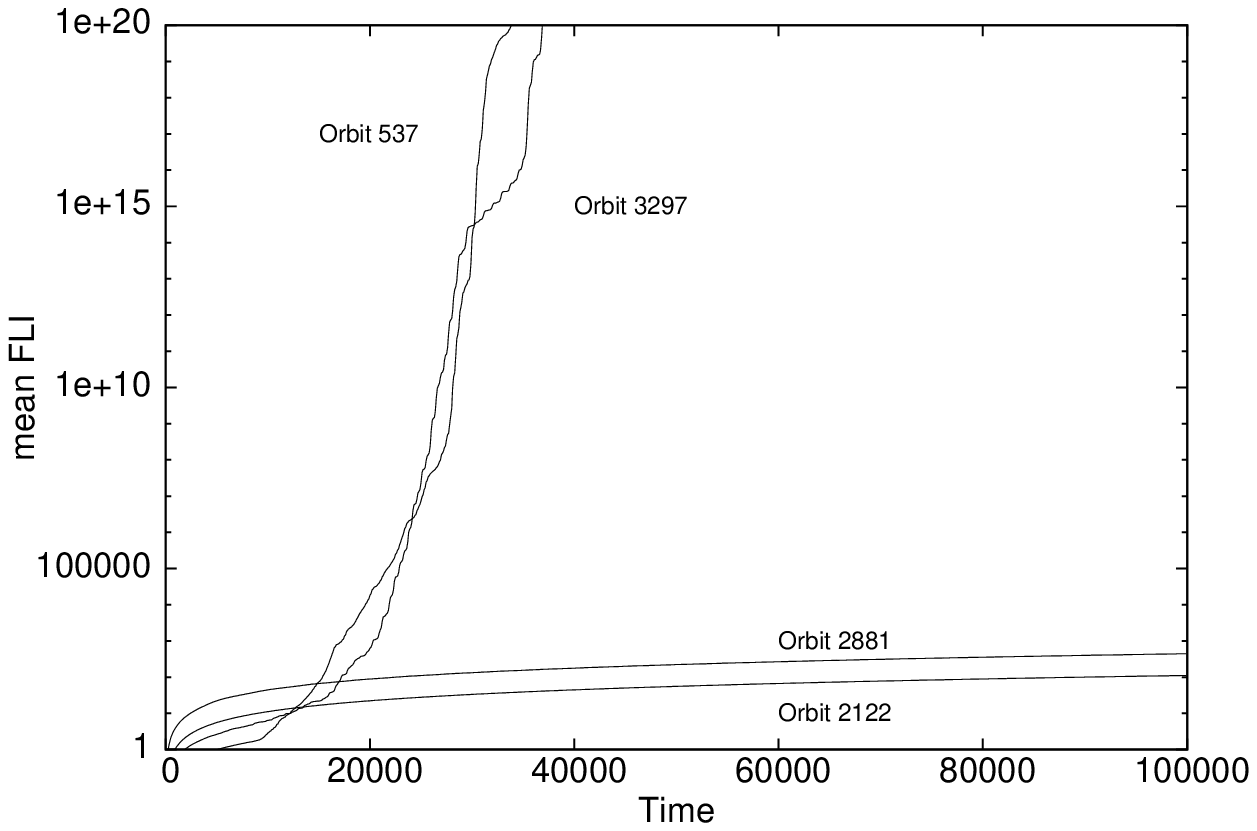}}
\end{tabular}
\caption{Time evolution of the MEGNO, the LCN and
the mean FLI for the sample orbits in
$\mathbf{Or}_d^u$ and $\mathbf{Or}_d^s$ for $T=10000$ and $T=100000$ u.t. 
In the bottom panels we separately plot the evolution of the mean FLI for
different times.}
\label{fig5.1}
\end{center}
\end{figure}

Fig.~\ref{fig5.1} displays the temporal evolution of the three indicators for both
 integration times, namely, $T=10000$ and $T=100000$ u.t., corresponding to the selected sample orbits. 
It can clearly be observed that for $T=10000$ u.t. 
almost all orbits exhibit a stable behavior. Yet, both the MEGNO and the mean FLI evince an incipient 
increase for orbits 537 and 3297, which is missed by the LCN.  
As time increases, both orbits clearly separate from the rest, all the three indicators giving account 
of this fact.

Meanwhile, orbits
2122 $\in\mathbf{Or}_d^s$ and 2881 $\in\mathbf{Or}_d^u$ seem to evolve in a similar fashion. Though, 
the final MEGNO value for orbit 2122 is sharply $2$, while for  
2881 is slightly above the regular value,  which might 
indicate a rather mild unstable character of this orbit (see discussion below).


\begin{figure}[ht!]
\begin{center}
\begin{tabular}{cc}
\hspace{-5mm}\resizebox{63mm}{!}{\includegraphics{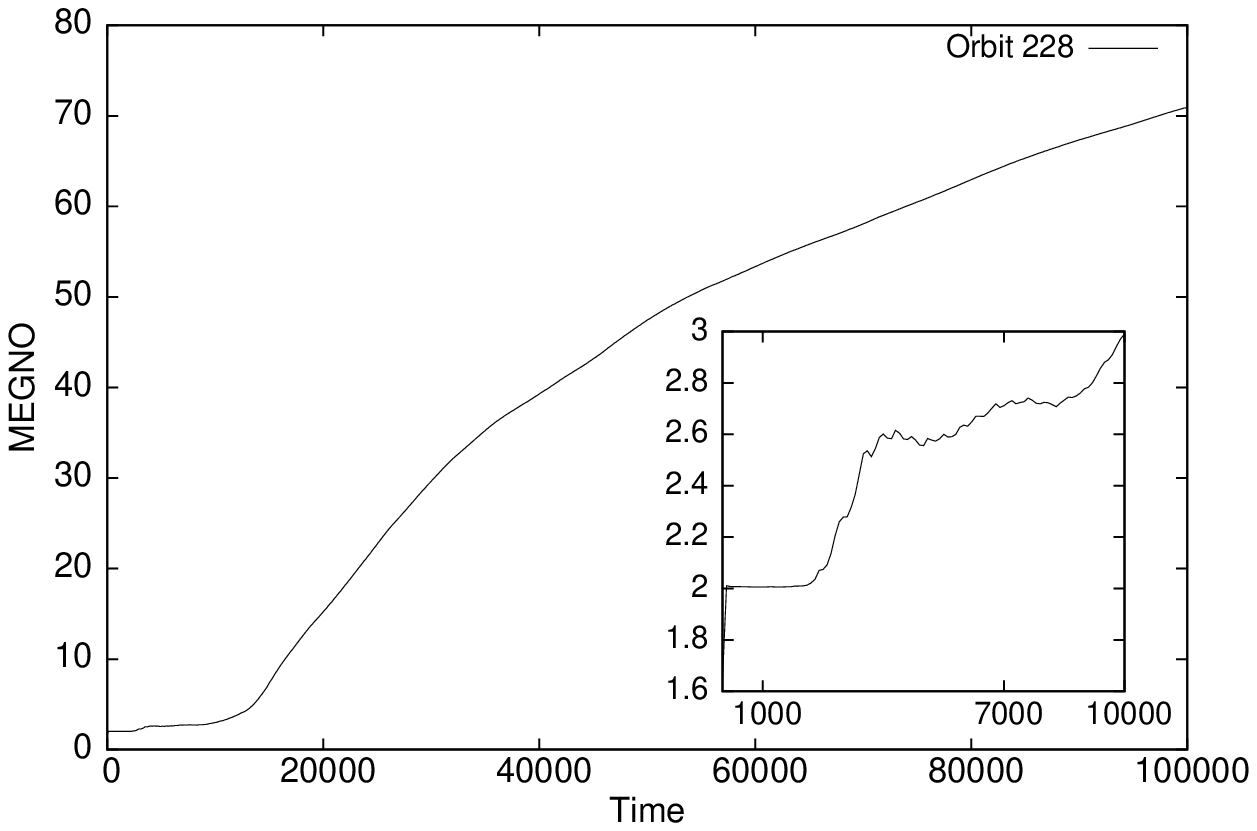}}&
\hspace{-5mm}\resizebox{63mm}{!}{\includegraphics{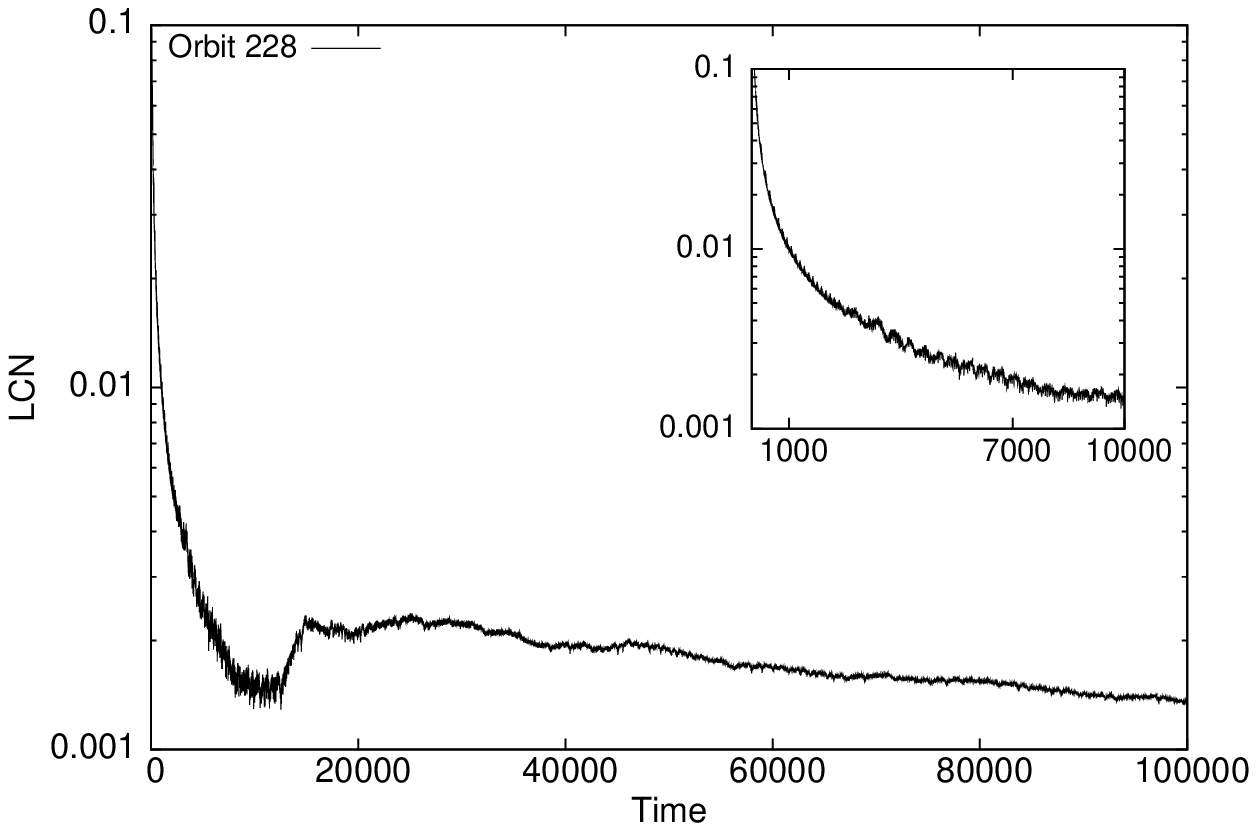}}
\end{tabular}
\caption{Time evolution of the MEGNO and the LCN for orbit 228 $\in\mathbf{Or}_d^u$ for
$T=10000$ and $T=100000$ u.t.}
\label{figure 5.2}
\end{center}
\end{figure}

In Fig~\ref{figure 5.2} we present the time evolution of both the MEGNO and the LCN, 
on the left and right panel respectively, for orbit 228 $\in\mathbf{Or}_d^u$. It
is interesting to note the particular behavior of the indicators for this orbit. 
From the plot on the left, the trajectory  looks like a stable quasi-periodic orbit up to $t\lesssim 2000$ u.t., then the MEGNO grows linearly 
for a rather short time interval to  reach a nearly constant value, around
$2.6$, and attains the value $3$ at $T=10000$ u.t. Note that the LCN at this time is very
close to the theoretical expected one, around $0.001$. When  the integration time is increased 
the MEGNO grows up to higher values but not in a linear fashion, while the LCN seems to 
 decrease, though it approaches a larger value than the one corresponding to regular
motion at $T=100000$ u.t.  The peculiar behavior of the indicators for this
orbit encourages  a more detailed study of its neigbourhood in phase space in order to grasp 
its actual dynamical nature.  
This chore will be performed by analysing 
its immediate neihgbourhood in phase space, which might provide us with valuable dynamical information.

\begin{figure}[ht!]
\begin{center}
\begin{tabular}{cc}
\hspace{-5mm}\resizebox{63mm}{!}{\includegraphics{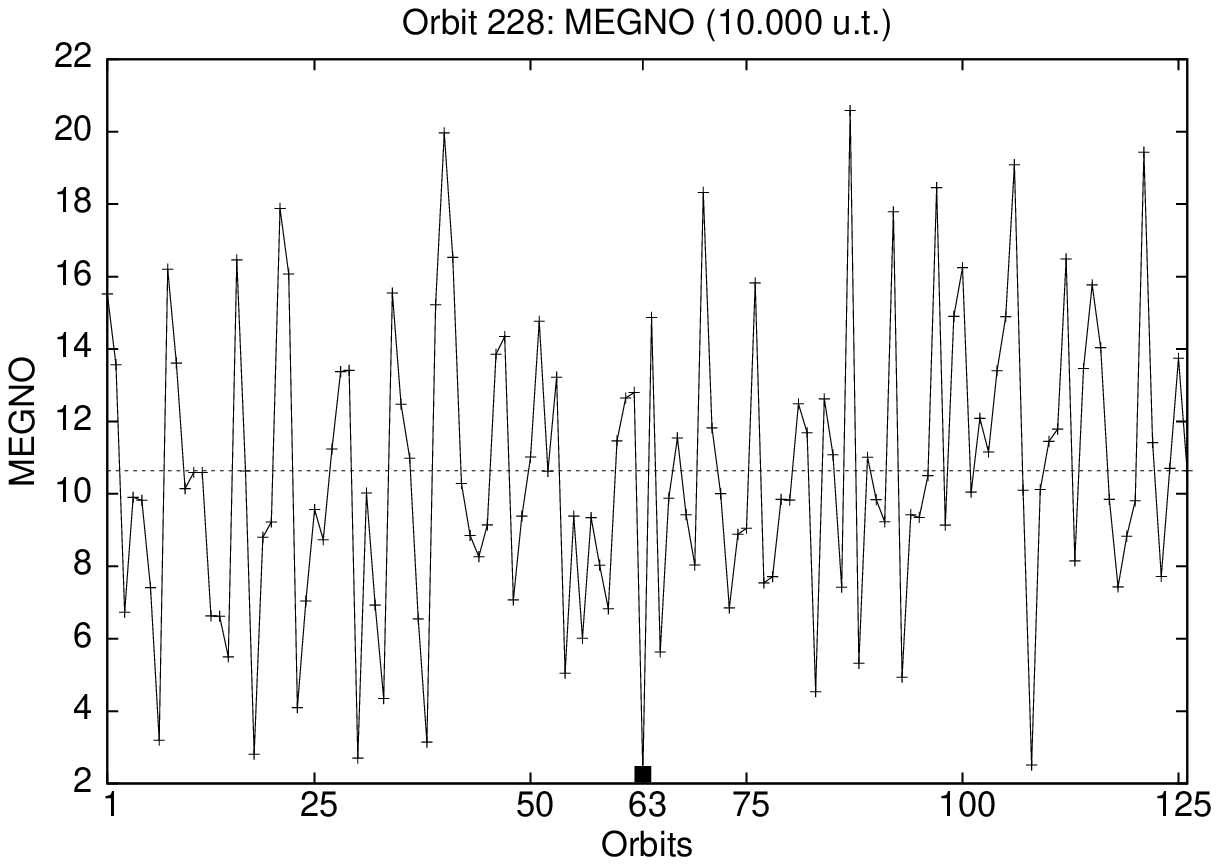}}&
\hspace{-5mm}\resizebox{63mm}{!}{\includegraphics{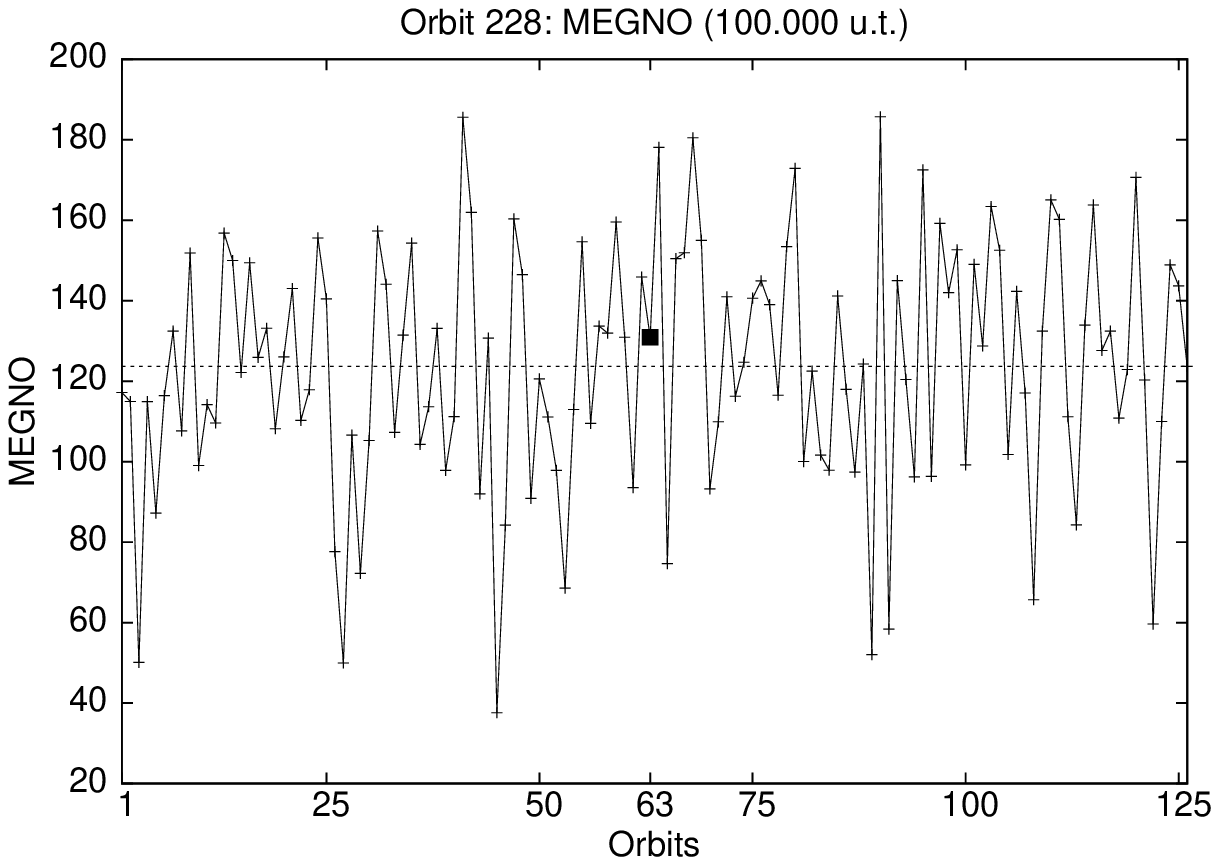}}
\end{tabular}
\begin{tabular}{cc}
\hspace{-5mm}\resizebox{63mm}{!}{\includegraphics{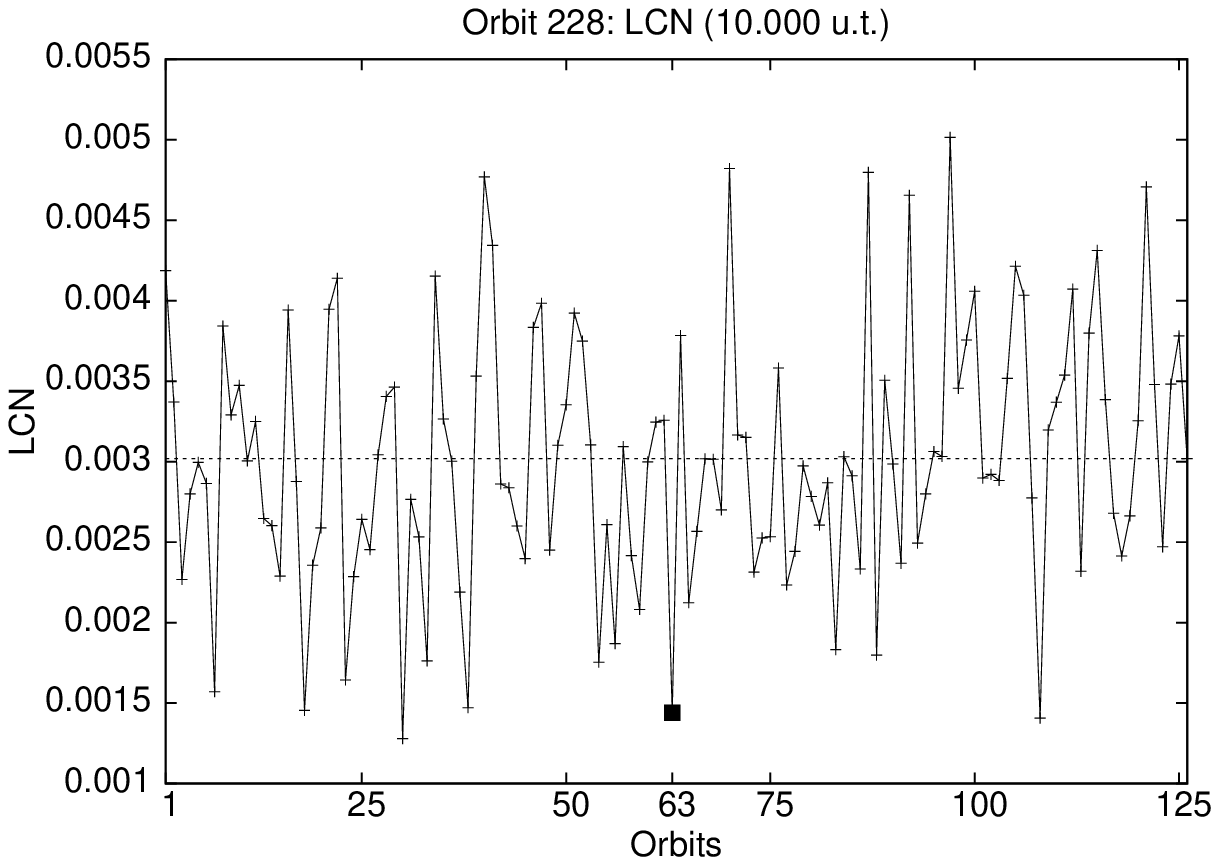}}&
\hspace{-5mm}\resizebox{63mm}{!}{\includegraphics{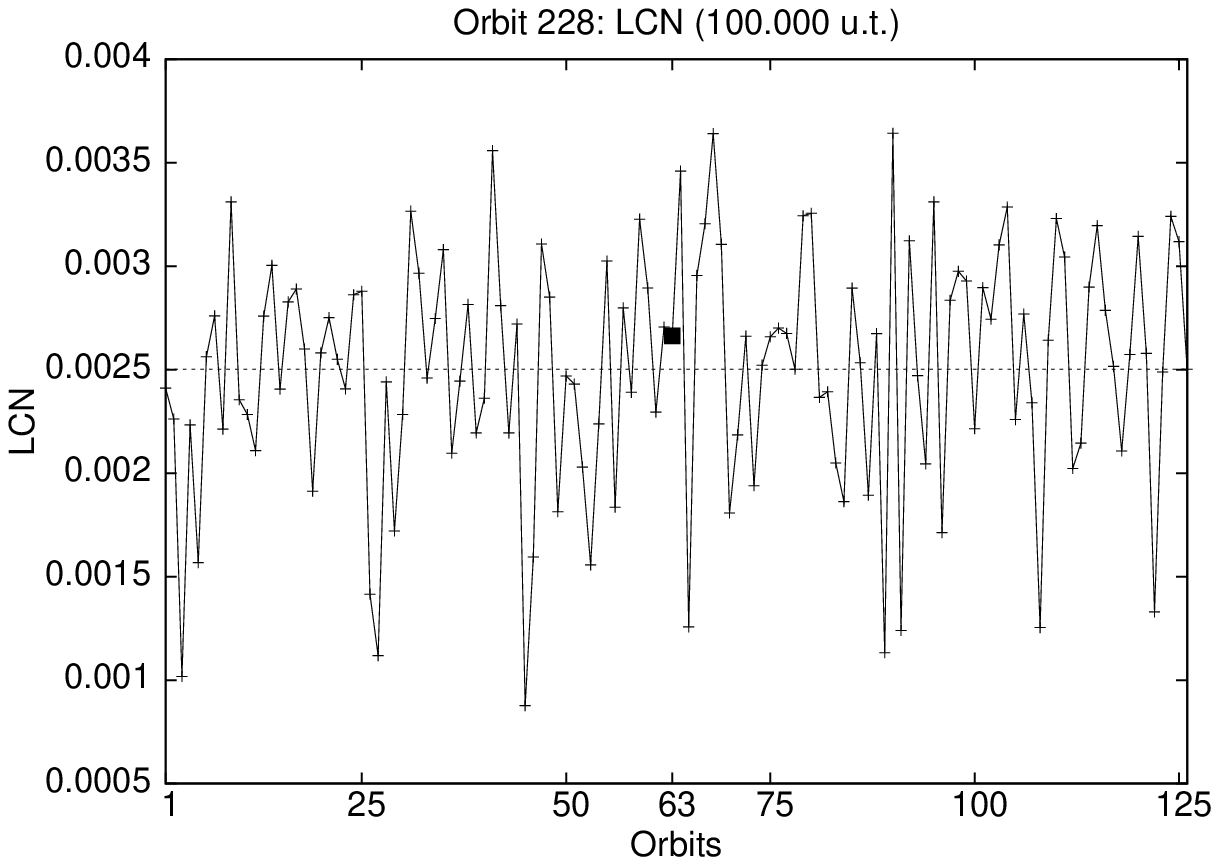}}
\end{tabular}
\caption{MEGNO and the LCN for $T=10000$ and $T=100000$ u.t. for 125
orbits in a domain of size $10^{-7}$ around orbit 228. The last value corresponds
to the mean value of the indicators, also represented by the horizontal line.}
\label{fig5.3}
\end{center}
\end{figure}

Thus, in Fig.~\ref{fig5.3} we present the MEGNO and LCN values at $T=10000$ and $T=100000$ u.t. for
a set of 125 orbits taken at random in a neighborhood of size $10^{-7}$ centered at orbit 228,  
whose concomitant values are depicted by full dots in each plot. 
Note that at $T=10000$ the MEGNO for this 
orbit is very close to 2, while the mean value of the indicator for this set of orbits 
 is about 11. 
On increasing the integration time, it becomes quite clear that orbit 228 is in fact chaotic. 
A similar behavior is observed in regards to the LCN values.  
Altogether, the figure 
suggests that this orbit might lie in a complex dynamical region of phase space.  

\begin{figure}[ht!]
\begin{center}
\resizebox{90mm}{!}{\includegraphics{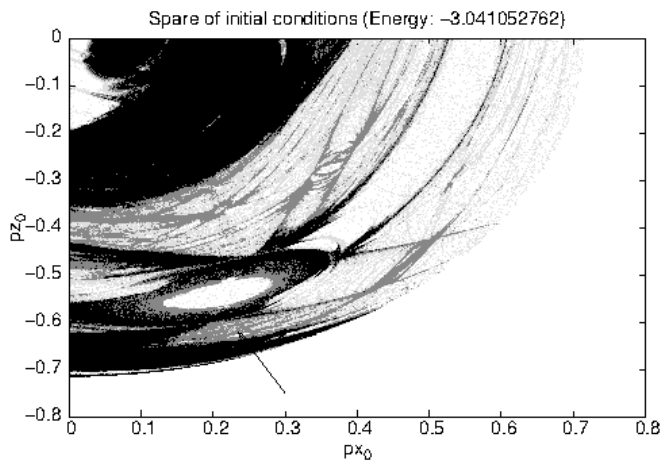}}
\caption{MEGNO contour plot for a grid of $10^6$ initial conditions in momenta space
$(p_x,p_z)$ for $E=-3.041052762$ and initial position of orbit 228. The total motion
time is $T=1000\approx 10^3T_c(E)$. Black ($\overline{Y}>20$) corresponds to
strong chaotic zones, white indicates regular regions $\overline{Y}<2.01$, light gray
to slight unstable or even some regular orbits ($2.01<\overline{Y}<3$) and dark gray indicates
mild chaotic domains ($3<\overline{Y}<20$).The arrow indicates the location of orbit 228.} 
\label{figure 5.4}
\end{center}
\end{figure}

In order to confirm this 
conjecture, in Fig.~\ref{figure 5.4} we show a MEGNO contour plot in the momenta space for 
the exact energy value of  orbit 228, constructed by taking as initial conditions the position of the orbit in configuration 
space and  $(p_x,p_z)$ varying over a grid of about $10^6$ points. 
The MEGNO values correspond to $T=1000$ u.t., which turns out 
to be a proper final time of integration since $T\approx 10^3T_c(E)$, and  
the characteristic time scale is about $1$ for an energy $E\sim -3$ (as follows from Fig.~\ref{figure 4.3}).

The dark regions corresponds
to strong chaotic motion for which $\overline{Y}>20$, while the white ones with
$\overline{Y}<2.01$ reveal stable motion. Light gray zones refer
to slight unstable (or even regular) motion, $2.01<\overline{Y}<3$ and the dark gray regions
represent mild chaotic motion $3<\overline{Y}<20$. This plot reveals
the complex resonance structure of phase space when projected onto the plane
$(p_x,p_z)$ at this energy level, and we can clearly 
see that orbit 228 lies inside a resonance crossing. 
This should explain the pathological behavior of this orbit and its surroundings.

\begin{figure}[ht!]
\begin{center}
\begin{tabular}{cc}
\hspace{-5mm}\resizebox{63mm}{!}{\includegraphics{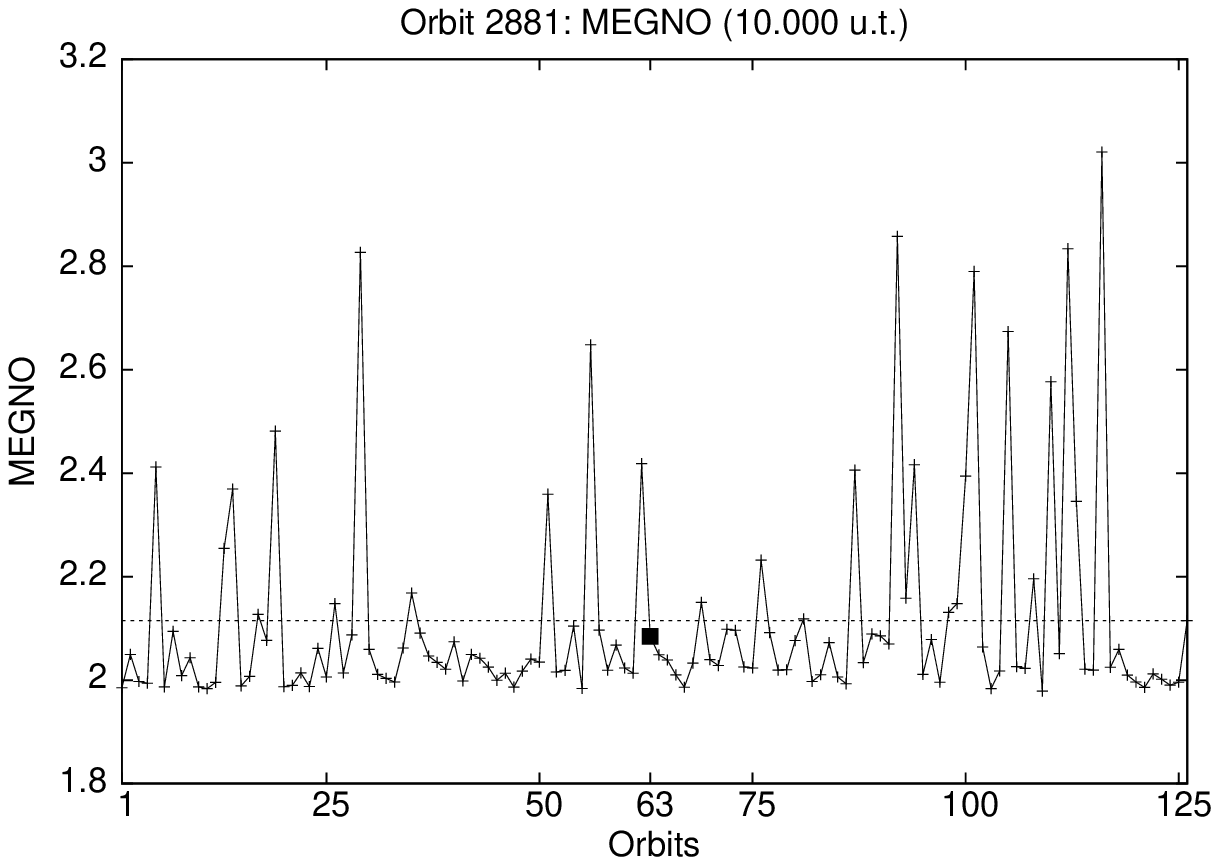}}&
\hspace{-5mm}\resizebox{63mm}{!}{\includegraphics{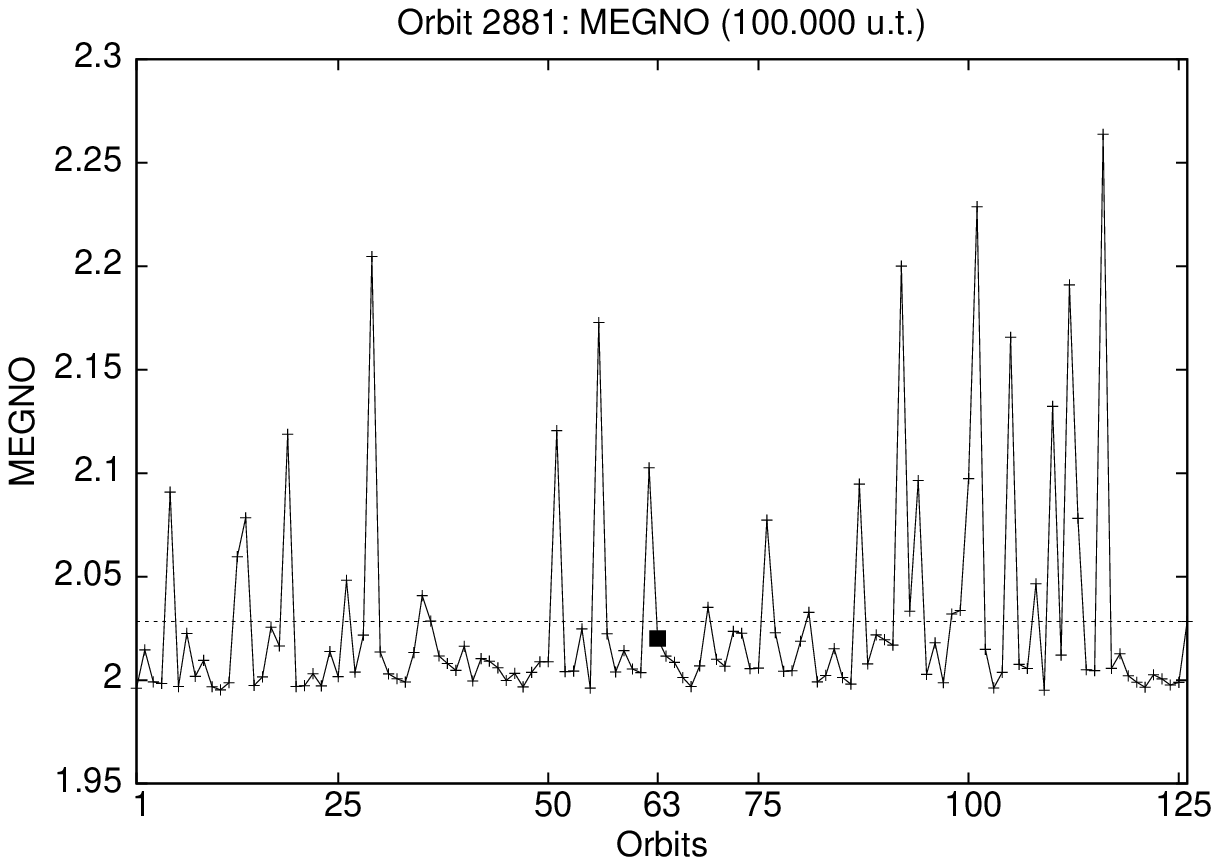}}
\end{tabular}
\begin{tabular}{cc}
\hspace{-5mm}\resizebox{63mm}{!}{\includegraphics{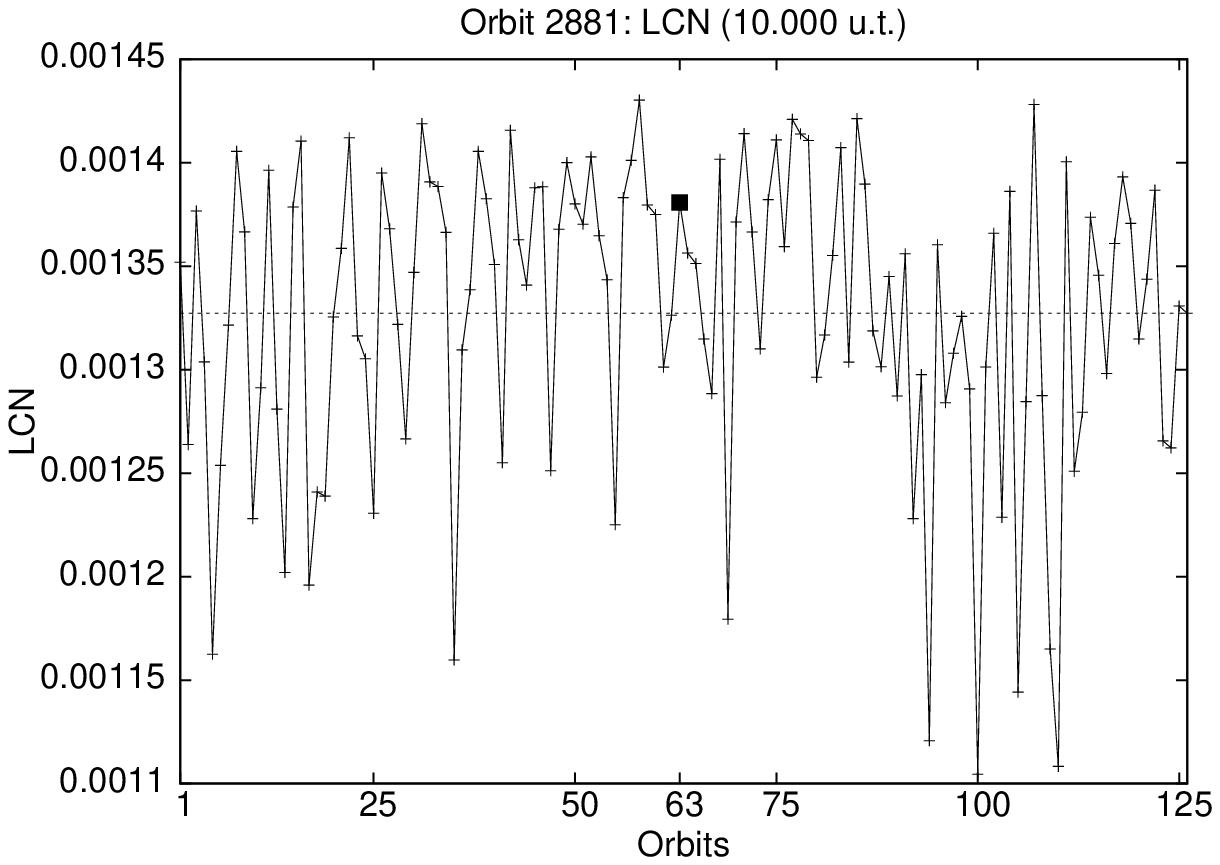}}&
\hspace{-5mm}\resizebox{63mm}{!}{\includegraphics{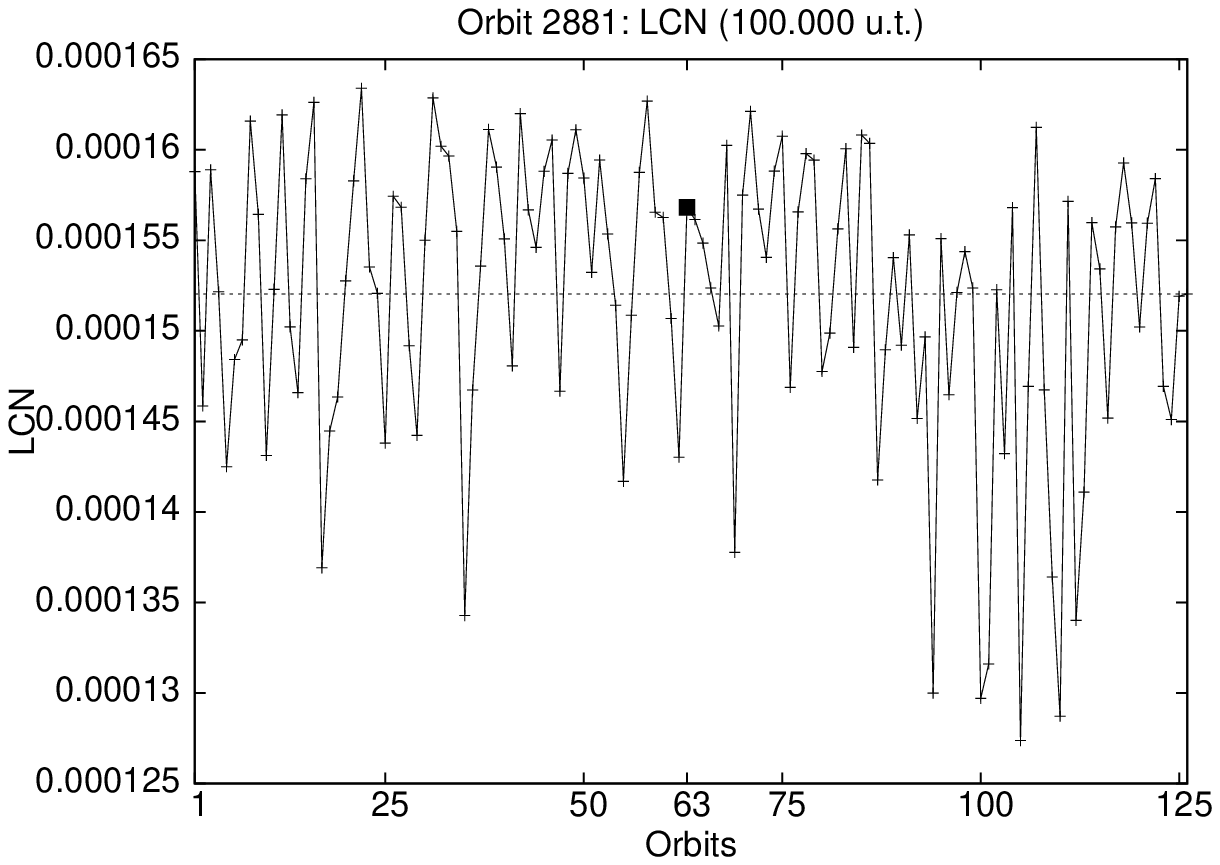}}
\end{tabular}
\caption{MEGNO and the LCN for $T=10000$ and $T=100000$ u.t. for 125
orbits in a domain of size $10^{-7}$ around orbit 2881. The last value corresponds
to the mean value of the indicators, also represented by the horizontal line.}
\label{fig5.5}
\end{center}
\end{figure}

Just to end this section, let us investigate the surroundings of orbit 2881
which belongs to $\mathbf{Or}_d^u$. 
For that sake, let us consider 125 orbits selected at random within a neibourhood of size $10^{-7}$ 
and compute both their MEGNO and LCN at $T=10000$ and $T=100000$ u.t. 
The results are displayed in Fig.~\ref{fig5.5} to show that the 
MEGNO values lie in the range $1.95\lesssim\overline{Y}\lesssim 3$ for $T=10000$, and
on considering larger motion times the MEGNO interval gets even narrower, e.g. $(1.98,2.27)$ 
for $T=100000$ u.t.
Therefore, though the MEGNO for orbit 2881 is slightly higher than the threshold $2.01$,
namely,  $2.014$ at $T=100000$, the orbit should be considered at all means stable. 
An analogous result provides the LCN, that has been computed 
for the very same orbits, including orbit 2881 for which, at $T=10000$ u.t., attains 
a value less than the empirical critical value $V_c^e$ adopted by \cite{MCW05}, but higher than $V_c^t$.

\begin{figure}[t!]
\begin{center}
\begin{tabular}{cc}
\hspace{-5mm}\resizebox{63mm}{!}{\includegraphics{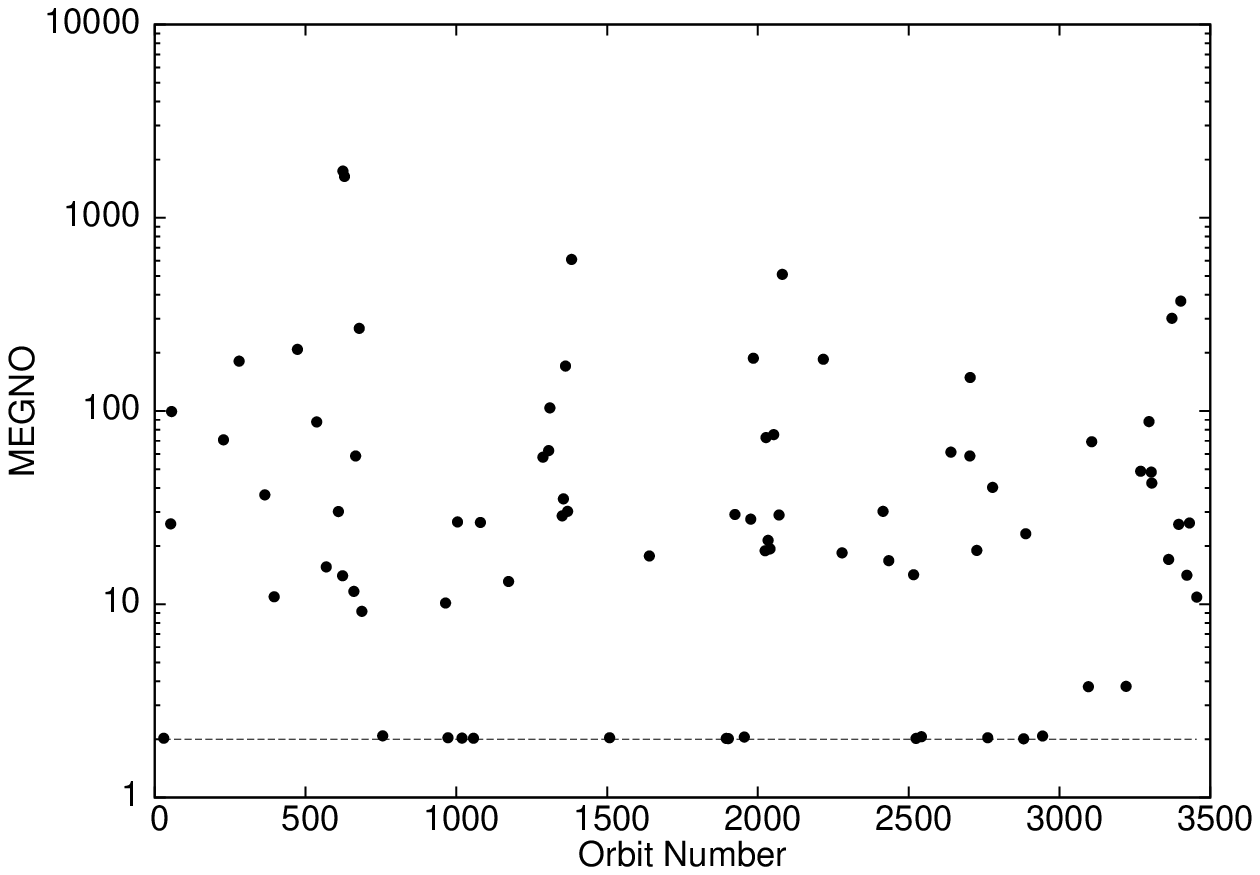}}&
\hspace{-5mm}\resizebox{63mm}{!}{\includegraphics{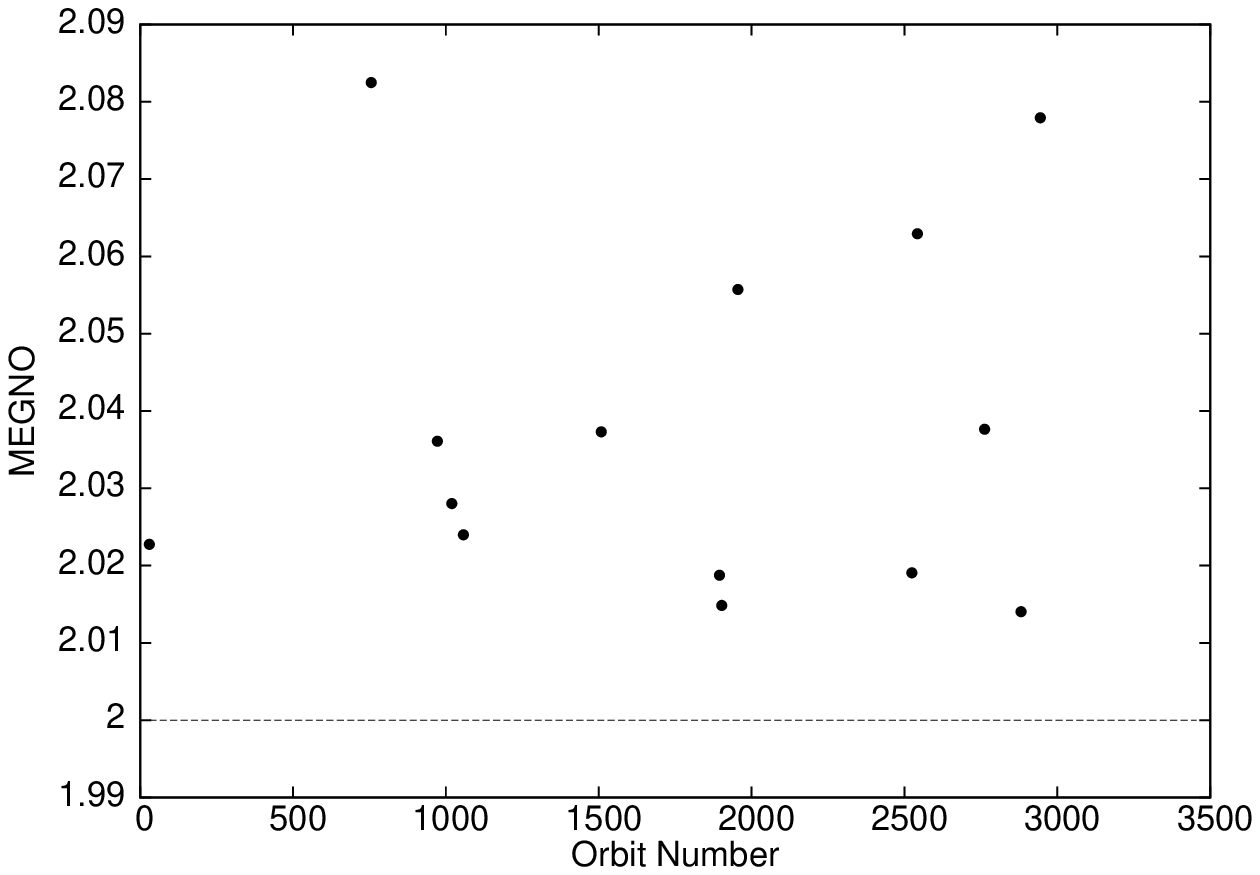}}
\end{tabular}
\caption{MEGNO values for the 79 orbits in $\mathbf{Or}_d^u$ for $T=100000$~(left). Zoom around
$\overline{Y}=2$~(right).}
\label{figure 5.6}
\end{center}
\end{figure}

Finally, it would be interesting to consider the MEGNO values of all orbits in
 $\mathbf{Or}_d^u$ for $T=100000$ u.t., which are displayed in Fig.~\ref{figure 5.6}.
It can there be noticed that at least 14 orbits in  $\mathbf{Or}_d^u$
should actually be included in the set $\mathbf{Or}_d^s$ (e.g. 2881), 
since their MEGNO values are rather too close to the regular value $2$. 
In fact, it might be inaccurate for them to consider the factor $2$ in 
 $2\overline{Y}/T$, necessary in the case of chaotic orbits, on looking for correlations with the LCN, 
since these orbits do not increase linearly with time. 

On transfering these $14$ orbits from $\mathbf{Or}_d^u$ to $\mathbf{Or}_d^s$, the recomputation of  
 the concomitant resulting distributions deliver for $\mathbf{Or}_d^u$, the mean value of 
$\log(2\overline{Y}/T)\approx-3.04$ with a standard deviation of about $0.55$ and the
mean of $\log(\mathrm{LCN})\approx-2.98$ with a standard deviation close to $0.51$, 
the correlation coefficient being $r\approx0.99$. For $\mathbf{Or}_d^s$ there results 
a mean value of $\overline{Y}/T\approx 2\times 10^{-5}$ with a standard deviation of $\approx2.3\times 10^{-7}$,
while the mean LCN  is $\approx 1.4\times 10^{-4}$ with a standard deviation of  
$\approx 1.8\times 10^{-5}$. The correlation coefficient in this case barely amounts $0.22$, indicating
again no correlation between both indicators for regular orbits.
Let us notice that no significant changes arise as a consequence of the transposition performed.

\section{Discussion}
\label{S5}

We have shown that the MEGNO is a suitable fast  indicator to separate  
regular from chaotic motion.  Further, it is particularly useful to investigate the nature
of orbits that have a small but positive Lyapunov number. 

Besides we have shown a rather good correlation between the MEGNO and the FT--LCN values for short, moderate
and large integration times for chaotic orbits, while the MEGNO provides better results
for regular motion. In fact, it has the advantage that $\overline{Y}/T$ converges to the null value
of $\sigma$ faster than the classical algorithm to compute the LCN.  Another recourse to derive
low values for the LCN in the case of quasi-periodic motion consists in computing the slope
of the MEGNO.

The FLI looks also as a reliable fast indicator, but it does not provide any reference
value for regular motion, so it may be useful to explore  phase space rather
than to investigate the nature of a given orbit, unless  we follow the time
evolution of this indicator.

Finally, though $\overline{Y}\to 2$ when $t\to\infty$ and for quasi-periodic motion
its convergence would be very fast, one should keep in mind that a single orbit is
in general dominated by the dynamics of its surroundings.
Therefore if it is necessary to determine the strict character of that orbit, 
the threshold value of a regular orbit would be $\overline{Y}\sim d$ where 
$d$ could be taken as the mean value of $\overline{Y}$ over a rather small domain
around the orbit.

\section*{Acknowledgments}
The authors are grateful to two anonymous referees for a careful reading of the 
manuscript and helpful recommendations.
This work was supported with grants from the Consejo Nacional de
Investigaciones Cient\'{\i}ficas y T\'ecnicas de la Rep\'ublica Argentina (CCT--La Plata),
the Agencia Nacional de Promoci\'on Cient\'{\i}fica y Tecnol\'ogica and
the Universidad Nacional de La Plata.





\bibliographystyle{elsarticle-num}
\bibliography{<your-bib-database>}

\begin{thebibliography}{00}
\bibitem{MCW05} Muzzio, J.C., Carpintero, D.D. and Wachlin, F.C., 2005, CeMDA, 91(1-2), 173
\bibitem{S10} Skokos, C., 2010, LNP, 790, 63S
\bibitem{CS00} Cincotta, P.M. and Sim\'o, C., 2000, A\&AS, 147, 205
\bibitem{CGS03} Cincotta, P.M., Giordano, C.M. and Sim\'o, C., 2003, Phys. D, 182, 151
\bibitem{G02} Go\'zdziewski, K., 2002,  A\&A, 393, 997
\bibitem{G03a} Go\'zdziewski, K., 2003a,  A\&A, 398, 315
\bibitem{G03b} Go\'zdziewski, K., 2003b,  A\&A, 398, 1151
\bibitem{PM03} Pavlov, A.I. and Maciejewski, A.J., 2003, ASTL, 29, 552
\bibitem{BKRP03} Bois, E., Kiseleva-Eggleton, L., Rambaux, N. and Pilat-Lohinger, E., 2003, ApJ, 598(2), 1312
\bibitem{G04} Go\'zdziewski, K. and Konacki, M., 2004, ApJ, 610(2), 1093
\bibitem{GKW05} Go\'zdziewski, K., Konacki, M. and Wolszczan, A., 2005, ApJ, 619(2), 1084
\bibitem{BMBW05} Breiter, S., Melendo, B., Bartczak, P. and Wytrzyszczak, I., 2005, A\&A, 437(2), 753
\bibitem{HMJGM08} Hinse, T.C., Michelsen, R., J\o rgensen, U.G., Go\'zdziewski, K. and Mikkola, S., 2008, A\&A, 488(3), 1133
\bibitem{GB08} Gayon, J. and Bois, E., 2008, A\&A, 482(2), 665
\bibitem{LDV09} Lema\'{\i}tre, A., Delsate, N. and Valk, S., 2009, CeMDA, 104, 338L
\bibitem{BBS09} Barrio, R., Blesa, F., Serrano, S., 2009, Phys. D, 238, 1087
\bibitem{HCA09} Hinse, T., Christou, A.; Alvarellos, J., 2009arXiv0907.4886H
\bibitem{VDLC09} Valk, S.; Delsate, N.; Lema\^itre, A.; Carletti, T., AdSpR, 2009, 43, 10, 1059
\bibitem{GC04} Giordano, C.M. and Cincotta, P.M., 2004, A\&A, 423, 745
\bibitem{FGL97} Froeschl\'e, Cl., Gonczi, R. and Lega, E., 1997, P\&SS, 45, 881
\bibitem{CV97} Contopoulos, G and Voglis, N., 1997, A\&A, 317, 73 
\bibitem{S01} Skokos, C., 2001, Journal of Physics A, 34, 10029
\bibitem{SBE00} S\'andor, Z., B\'alint, \'E. and Efthymiopoulos, C., 2000, CeMDA, 78, 113 
\bibitem{LVE08} Lukes-Gerakopoulos, G., Voglis, N. and Efthymiopoulos, C., 2008, Physica A, 387, 1907
\bibitem{BS82}  Binney, J and Spergel, D., 1982, ApJ, 252, 308
\bibitem{L90} Laskar, J., 1990, Icarus, 88, 266
\bibitem{L93} Laskar, J., 1993, Phys. D, 67, 257
\bibitem{SM96} Sidlichovsk\'y, M. and Nesvorn\'y, D., 1996, CeMDA, 65, 137
\bibitem{CO08} Cordani, B, 2008, Phys. D,  237, 2797
\bibitem{VKS02} Voglis, N., Kalapotharakos, C. and Stavropoulos, L., 2002, MNRAS, 337(2), 619
\bibitem{M06} Muzzio, J.C., 2006, CeMDA, 96(2), 85
\bibitem{CGM08} Cincotta, P.M., Giordano, C.M. and Muzzio, J.C., 2008, Discrete and Continuos Dynamical Systems B, 10, 439
\bibitem{HNW87} Hairer,~E., N\o rsett,~S. and Wanner,~G., 1987, Solving Ordinary Differential Equations I: Nonstiff Problems, Springer--Verlag
\bibitem{PD81} Prince,~P. and Dormand,~J., 1981, J. Comp. Appl. Math., 35, 67
\bibitem{Wessa09} Wessa, P. (2009), Free Statistics Software, Office for Research Development and Education, version 1.1.23-r4, URL http://www.wessa.net/
\bibitem{BGS76} Benettin, G., Galgani L., and Strelcyn, J. M., 1976, Phys. Rev. A, 14(6), 2338
\end{thebibliography}







\end{document}